\documentclass{aa}  

\usepackage{graphicx}
\usepackage{txfonts}

\usepackage{booktabs}

\usepackage[colorinlistoftodos,color=blue!20]{todonotes}

\usepackage{ulem}
\usepackage{multirow}
\usepackage{longtable}
\usepackage{hyperref}
\hypersetup{pdftex, colorlinks=true, linkcolor=blue, citecolor=blue, filecolor=blue, urlcolor=blue, pdftitle= Astrophysics”; http://www, pdfauthor=E389, pdfsubject=, pdfkeywords=}

   \title{Molecular line study of the S-type AGB star W\,Aquilae}

   \subtitle{ALMA observations of CS, SiS, SiO and HCN}

   \author{M.~Brunner
          \inst{1}
          \and
          T.~Danilovich
          \inst{2,3}
          \and
          S.~Ramstedt
          \inst{4}
          \and
          I.~Marti-Vidal
          \inst{2}
          \and
          E.~De~Beck
          \inst {2}
          \and
          W.H.T.~Vlemmings
          \inst{2}
          \and
          M.~Lindqvist
          \inst{2}
          \and
          F.~Kerschbaum
          \inst{1}
          }

   \institute{Department for Astrophysics, University of Vienna,
              T\"urkenschanzstrasse 17, A-1180 Vienna\\
              \email{magdalena.brunner@univie.ac.at}
         \and
          Department of Space, Earth and Environment, Chalmers University of Technology, Onsala Space Observatory, 439 92 Onsala, Sweden
          \and
          Department of Physics and Astronomy, Institute of Astronomy, KU Leuven, Celestijnenlaan 200D,  3001 Leuven, Belgium
          \and
         Department of Physics and Astronomy, Uppsala University, 75120 Uppsala, Sweden }

   \date{Received Month nr, YYYY; accepted Month nr, YYYY}

  \abstract
   {With the outstanding spatial resolution and sensitivity of the Atacama Large Millimeter/sub-millimeter Array (ALMA), molecular gas other than the abundant CO can be observed and resolved in circumstellar envelopes (CSEs) around evolved stars, such as the binary S-type Asymptotic Giant Branch (AGB) star W~Aquilae.}
   {We aim to constrain the chemical composition of the CSE and determine the radial abundance distribution, the photospheric peak abundance, and isotopic ratios of a selection of chemically important molecular species in the innermost CSE of W~Aql. The derived parameters are put into the context of the chemical evolution of AGB stars and are compared with theoretical models.}
   {We employ one-dimensional radiative transfer modeling -- with the accelerated lambda iteration (ALI) radiative transfer code -- of the radial abundance distribution of a total of five molecular species (CS, SiS, $^{30}$SiS, $^{29}$SiO and H$^{13}$CN) and determine the best fitting model parameters based on high-resolution ALMA observations as well as archival single-dish observations. The additional advantage of the spatially resolved ALMA observations is that we can directly constrain the radial profile of the observed line transitions from the observations.}
   {We derive abundances and $e$-folding radii for CS, SiS, $^{30}$SiS, $^{29}$SiO and H$^{13}$CN and compare them to previous studies, which are based only on unresolved single-dish spectra. Our results are in line with previous results and are more accurate due to resolution of the emission regions.}
   {}

\usepackage{color}

\begin{document}

   \maketitle
%

\section{Introduction}

W\,Aquilae (\object{W Aql}) is a cool, intermediate-mass S-type star located on the Asymptotic Giant Branch (AGB). In this phase of stellar evolution, a star suffers from substantial mass loss, and, as a consequence, a circumstellar envelope (CSE) is created. In this circumstellar environment, molecules and dust grains are formed, making AGB stars important contributors to the chemical evolution of the interstellar medium and ultimately the universe \citep[e.g.][]{agbbook}. On the AGB, different molecular bands in near-infrared (NIR) spectra (e.g., TiO, ZrO, CN) are used to discriminate between three different spectral types of AGB stars. These are also thought to reflect their carbon-to-oxygen (C/O) ratios, measured in the stellar atmosphere: M-type (oxygen-rich, C/O < 1), S-type (C/O $\simeq$ 1) and C-type (carbon-rich, C/O > 1). Owing to the varying circumstellar chemical composition, certain molecular and dust species will be formed preferentially to others, depending on the chemical type of an AGB star. 

In AGB stars, carbon is produced from helium through the triple-alpha process. A deep convective envelope and dredge-up events mix the freshly created carbon up into the stellar atmosphere, where the C/O ratio can be measured. The dredged-up carbon increases the C/O ratio, leading to a change in the circumstellar chemistry and the evolution from M-type to C-type stars. This implies that S-type stars represent transition objects in the chemical stellar evolution on the AGB, connecting two very distinct chemical types of stars and stellar atmospheres. As reviewed in \citet{Hofner2015}, the stellar wind is most likely driven by radiation pressure on dust grains, meaning that the differing chemical composition of M- and C-type stars will require different wind-driving dust species. Up to now, the wind acceleration process and the respective wind-driving dust species are well studied for C-type stars and recently solutions have been suggested for M-type stars as well \citep[e.g.][]{Hofner2008, Bladh2012, Bladh2015}. Comparable studies for S-type stars are difficult, however, given the uncertain circumstellar chemistry. This makes the detailed chemical analysis of different types of AGB stars crucial for our understanding of this evolutionary stage.

For the determination of properties such as mass-loss rates or circumstellar morphology, the observation and analysis of the abundant and stable CO molecule is most commonly used, which is present in the CSEs of all three chemical types \citep[e.g.][]{Knapp1998, Schoier2002,  Olofsson2002, Ramstedt2009, De-Beck2010, Danilovich2015}. Otherwise, it is commonly observed that the CSEs of oxygen-rich stars are dominated by oxygen-bearing molecules, while the CSEs of carbon-rich stars are dominated by carbon-bearing molecules. Apart from that simplified picture, it was predicted by \citet{Cherchneff2006} and confirmed by \citet{Decin2008} that under non-LTE conditions a small selection of so-called ``parent'' molecules is present for all three chemical types, concentrated to the inner stellar wind. These ``parent'' molecules -- CO, SiO, HCN, CS -- can be formed in shock waves propagating through the outer layers of the stellar photosphere, and these molecules will influence the CSE chemistry and abundances. Recent observations suggest that the parent molecule inventory is supplemented with the species H$_{2}$O \citep{Maercker2016, Danilovich2014, Neufeld2014} and NH$_{3}$ \citep{Menten2010, Danilovich2014, Schmidt2016}. For S-type stars, the chemical composition is highly sensitive to the exact C/O ratio and temperature, leading to the splitting into MS-, S- , and SC-type stars for different molecular features seen in the spectra, with the C/O ratio of S-type stars ranging from 0.75 to 0.99 \citep{Zijlstra2004, Van-Eck2017}. The types MS and SC show similar molecular species to M- and C-type stars, while ``pure'' S-type stars show a mixed or intermediate molecular composition \citep[e.g.][]{Schoier2013}. Nevertheless, \citet{Ramstedt2009} derive that the mass-loss rate distributions of M-, S- and C-type AGB stars are indistinguishable, which makes it even more important to investigate how the different chemical compositions can lead to a similar outcome.\\

Observations, in combination with detailed radiative transfer modelling, are indispensable in the investigation of the chemical composition of AGB CSEs. Their study yields chemical abundance distributions and isotopolog ratios. One prominent example for an in-depth study of the molecular CSE is the carbon AGB star IRC+10216, which is probably the best analyzed AGB star so far, and has for example been observed within an interferometric spectral line survey with the Submillimeter Array (SMA) executed by \citet{Patel2011}. With this survey of unprecedented sensitivity, a total of 442 spectral lines, including more than 200 new ones, were detected. Although the emission has been mapped interferometrically, only a few of the strongest lines have been resolved.
As discussed by, for example, \citet{Saberi2017}, resolving the emitting region of molecules studied in this way strongly influences the derived abundance and radial size of the molecular envelopes, increasing the need for high-resolution observations of AGB CSEs and in particular their  inner regions. This has now become possible with the Atacama Large Millimeter/sub-millimeter Array (ALMA), which provides the necessary spatial and spectral resolution as well as the sensitivity to observe even weak molecular emission close to AGB stars.

In this paper, we present ALMA observations of the molecules CS, SiS, $^{29}$SiO and H$^{13}$CN, as well as four different isotopologs and vibrational states of SiO and SiS around the S-type AGB star W\,Aql. The star and previous studies of it are introduced in Sect.\,\ref{sec:waql}. The observations are described and presented in Sect.\,\ref{sec:observations}, while detailed maps of all molecular lines are shown in Appendix\,\ref{sec:appendix1}. The observational results are discussed in Sect.\,\ref{sec:obsresults}. Additionally, the sizes of the emitting regions are fitted in the uv-plane to further analyze the morphology of the molecular gas and partially unravel the three-dimensional (3D) structure of the CSE. Section\,\ref{sec:radtransfer} describes the radiative transfer modeling based on the work of \citet{Danilovich2014}, which is improved by the incorporation of the new ALMA data. The main uncertainty in circumstellar abundance estimates is the size of the emitting region. By measuring it directly with ALMA, this uncertainty is removed. The results of this whole study are summarized and discussed in Sect.\,\ref{sec:discussion}, and the main conclusions are given in Sect.\,\ref{sec:conclusions}.


\section{W\,Aql}
\label{sec:waql}
W\,Aql is an S-type AGB star with a spectral type of S6.6, which is very close to the SC-type classification, and has an estimated C/O ratio of 0.98 according to the classification scheme by \citet{Keenan1980}. It has a period of 490\,days and is classified as a Mira variable \citep{Feast2000}. Its distance is 395\,pc, derived from the period-magnitude relation \citep{Whitelock2008} and 2MASS $K$ band magnitude \citep{Cutri2003}. As reported by \citet{Decin2008}, the ``parent'' molecules CO, HCN, CS, SiO and SiS, expected to be formed close to the stellar photosphere, have all been detected for W\,Aql, but have not yet been resolved. The earlier study by \citet{Bieging1998} also suggests that SiO and HCN detected around W\,Aql could be of photospheric origin, under certain non-LTE conditions.

The chemical constitution of the CSE around W\,Aql has been investigated in several studies with observations and subsequent radiative transfer modeling \citep[e.g.,][]{Decin2008, De-Beck2010, Ramstedt2009, Ramstedt2014, Schoier2013, Danilovich2014}, but until now the observations of the molecular line emission have always been unresolved. According to \citet{Uttenthaler2013}, the star is Tc rich, confirming its evolutionary stage between M-type and carbon-rich AGB stars. Based on \citet{Danilovich2014}, the stellar effective temperature, $T_{eff}$ of W\,Aql is assumed to be 2300\,K, the stellar luminosity $L$ is 3000\,$L_{\odot}$ and the gas mass-loss rate is $4\times 10^{-6}\,M_{\odot} yr^{-1}$.

 \citet{Ramstedt2011} have shown W\,Aql to be a binary star, where they present an HST B-band image of the resolved binary system, with the binary located to the South-West. The angular separation of 0.46$^{\prime\prime\,}$ corresponds to a minimum projected binary separation of 190\,AU (for the derived distance of 395\,pc), making it a wide binary system. An optical spectroscopic classification of the binary companion was done by \citet{Danilovich2015} and lead to the spectral type F8 or F9, where the binary companion most likely lies behind or within the CSE of W\,Aql. Additionally, this analysis constrained the mass of the AGB star to 1.04 - 3\,M$_{\odot}$, leading to a total system mass of 2.1 - 4.1\,M$_{\odot}$. Newer mass estimates, based on observations of the $^{17}$O/$^{18}$O ratio and comparison with stellar evolution models by \citet{De-Nutte2017} give values of between 1.5 and 1.8\,M$_{\odot}$ for the initial mass of the primary star.

Asymmetries in the CSE of W~Aql have been detected at several different spatial scales and at different wavelengths. Interferometric observations of the dusty CSE of W\,Aql with the UC Berkeley Infrared Spatial Interferometer (ISI) at 11.15$\,\mu m$ were reported by \citet{Tatebe2006}, revealing an asymmetry of the dust distribution within a radius of 500\,milli-arcseconds. A dust excess was detected to the east of the star, appearing to continuously extend from the star to the outer regions of dust emission probed by the observations. This leads the authors to the conclusion that the asymmetry has been stable over the last 35\,years, which they estimate is the period since the ejection of the asymmetrically distributed material, using a typical wind velocity of 20\,km~s$^{-1}$. The authors argue that a close binary would disturb the stability of the asymmetry.

 At larger spatial scales, observations with the Polarimeter and Coronograph (PolCor) at the Nordic Optical Telescope (NOT) have been carried out by \citet{Ramstedt2011}, revealing the dust CSE morphology of W\,Aql on scales from about 1$^{\prime\prime\,}$ to 20$^{\prime\prime\,}$. They report a clear asymmetry to the south-west of the star, extending out to almost 10$^{\prime\prime\,}$, which is seen both in the total intensity image in the R band as well as in the polarised images, showing polarisation degree and angle. The direction of the asymmetry aligns with the binary position.

W\,Aql was observed with the PACS photometer of the Herschel space telescope as part of the MESS \citep[Mass loss of Evolved StarS; ][]{Groenewegen2011} program to analyze the large scale morphology of the dusty CSE and wind-ISM interaction \citep{Mayer2013} on spatial scales from approximately 10$^{\prime\prime\,}$ to 120$^{\prime\prime\,}$. The authors conclude that a strong feature in the 70$\,\mu m$ IR image of the CSE to the east of the star can be explained by the interaction of the stellar wind with an ISM flow. They further report the shaping of the IR emission by wind-binary interaction, which can be tentatively traced with an Archimedean spiral and could be responsible for additional brightening of the ISM interaction zone to the east of the star and the overall elliptical shape of the CSE.
The high-resolution CO gas emission around W\,Aql is presented and discussed in \citet{Ramstedt2017}, where arc-like structures at separations of approximately 10$^{\prime\prime\,}$ are found. While these structures can be linked to the binary nature of the object, small-scale arc structures at separations of 2-3$^{\prime\prime\,}$ remain unexplained by the known orbital parameters of the binary system. The small-scale structures are confined to the south-west region of emission, which is also consistent with the previously detected asymmetry found in dust emission.

\section{Observations and data reduction}
\label{sec:observations}
W\,Aql was observed as part of a sample of binary stars with ALMA in Cycle 1 (observed in March and April 2014, PI: Sofia Ramstedt), combining ALMA main array observations (12\,m diameter, long baselines), Atacama Compact Array (ACA) observations (7\,m diameter, short baselines) and total power (TP) observations (12\,m diameter, single-dish). An area of 25$^{\prime\prime\,}\times$ 25$^{\prime\prime\,}$ is covered with both arrays, observing a mosaic of 10 and 3 pointings, respectively. Four spectral windows, with a bandwidth of 1875\,MHz (main array) and 1992.187\,MHz (ACA) each, were observed, centered at 331, 333, 343 and 345\,GHz, respectively. The resulting spectral resolution is 488.281\,kHz. Standard calibration was performed with the Common Astronomy Software Application (CASA), using Ceres as flux calibrator and the quasars J1924-2914 and J1911-2006 as bandpass calibrator and phase calibrator, respectively. In addition to the main goal of the proposal, focussing on the analysis of the binary interaction on the CO gas distribution \citep{Ramstedt2017}, eight additional molecular lines were detected with this configuration. The molecular lines, rest frequencies and final velocity resolution after calibration and imaging are listed in Table\,\ref{tab:mollist}. 
The combination of the main array and ACA data is conducted in the visibility plane with antenna-specific weighting, whereas the TP data is combined with the rest of the data via feathering in the image plane. Imaging of all spectral lines was carried out with CASA using the CLEAN algorithm with natural weighting, resulting in a beam of 0.55$^{\prime\prime\,} \times$ 0.48$^{\prime\prime\,}$ (PA 86.35$^\circ$). Iterative masking for each spectral channel and molecular line was used to optimize the imaging process with CLEAN. The spectral channels were binned to a velocity resolution of 1\,km~s$^{-1}$ and 2\,km~s$^{-1}$ for strong and weak spectral lines, respectively. \\

\begin{table*}[htp]
\caption{Observed quantities of the molecular line emission around W~Aql.}
\begin{center}
\begin{tabular}{lcccccc}
\toprule
\multicolumn{1}{l}{Molecular line}
& \multicolumn{1}{l}{\parbox{2cm}{\centering Restfreq. \\(GHz)}}
& \multicolumn{1}{l}{\parbox{2cm}{\centering $\Delta$v \\(\,km~s$^{-1}$)}}
& \multicolumn{1}{l}{\parbox{2cm}{\centering $I_{\nu,peak}$ \\(Jy)}}
& \multicolumn{1}{l}{\parbox{2cm}{\centering $v_{lsr,peak}$ \\(\,km~s$^{-1}$)}}
& \multicolumn{1}{l}{\parbox{2cm}{\centering $\int I_{\nu} \, d\nu$ \\(Jy \,km~s$^{-1}$)}}
& \multicolumn{1}{l}{\parbox{2cm}{\centering $\theta$ \\($^{\prime\prime\,}$)}}
\\

\midrule
CS\,(7--6)& 342.8829 & 1 & 7.29 & -13 & 167.77 & 4 \\
H$^{13}$CN\,(4--3) & 345.3398 & 1 & 3.90 & -21 & 96.73 & 4\\ 
SiS\,(19--18) & 344.7795 & 1 & 3.51 & -15 & 70.51 & 4\\
$^{29}$SiO\,(8--7) & 342.9808 & 1 & 2.91 & -22 & 54.36 & 4\\
\addlinespace
SiO\,v=1\,(8--7) & 344.9162 & 2 & 0.74 & -23 & 11.33 & 4\\
SiS\,v=1\,(19--18) & 343.1010 & 2 & 0.34 & -23 & 3.68 & 4 \\
$^{30}$SiS\,(19--18) & 332.5503 & 2 & 0.29 & -27 & 3.25 & 2\\
SiO\,v=2\,(8--7) & 342.5044  & 2 & 0.18 & -25 & 0.34 & 2 \\

\bottomrule
\end{tabular}
\end{center}
\tablefoot{Rest frequencies, velocity resolution ($\Delta$v), peak flux ($I_{\nu,peak}$), peak flux velocity ($v_{lsr,peak}$), integrated flux ($\int I_{\nu} \, d\nu$) and used aperture diameter ($\theta$) for all eight detected molecular lines around W\,Aql. The rms noise measured from line-free channels is 10\,mJy/beam.}
\label{tab:mollist}
\end{table*}%

\section{Observational results}
\label{sec:obsresults}

Four of the eight detected spectral lines are considerably weaker than the others, making analysis and especially modeling of these lines very uncertain or even impossible. Therefore, we concentrate our efforts on the strong spectral lines and present the analysis and modeling process for the strong lines in the following sections. Additionally, we refer to Appendix\,\ref{sec:appendix2} for a graphical representation of basic results on the weak lines. Channel maps of all strong spectral lines can be found in Appendix\,\ref{sec:appendix1}. Figures\,\ref{fig:strongmoments} and \ref{fig:weakmoments} show the integrated intensity (moment 0) maps for all observed spectral lines. A comparison between the spectra of the four strongest lines in this study is presented in Fig.\,\ref{fig:strongspec}.

       \begin{figure*}
   \centering
      \includegraphics[trim={7.5cm 0cm 7.5cm 3cm},clip,width=0.98\textwidth]{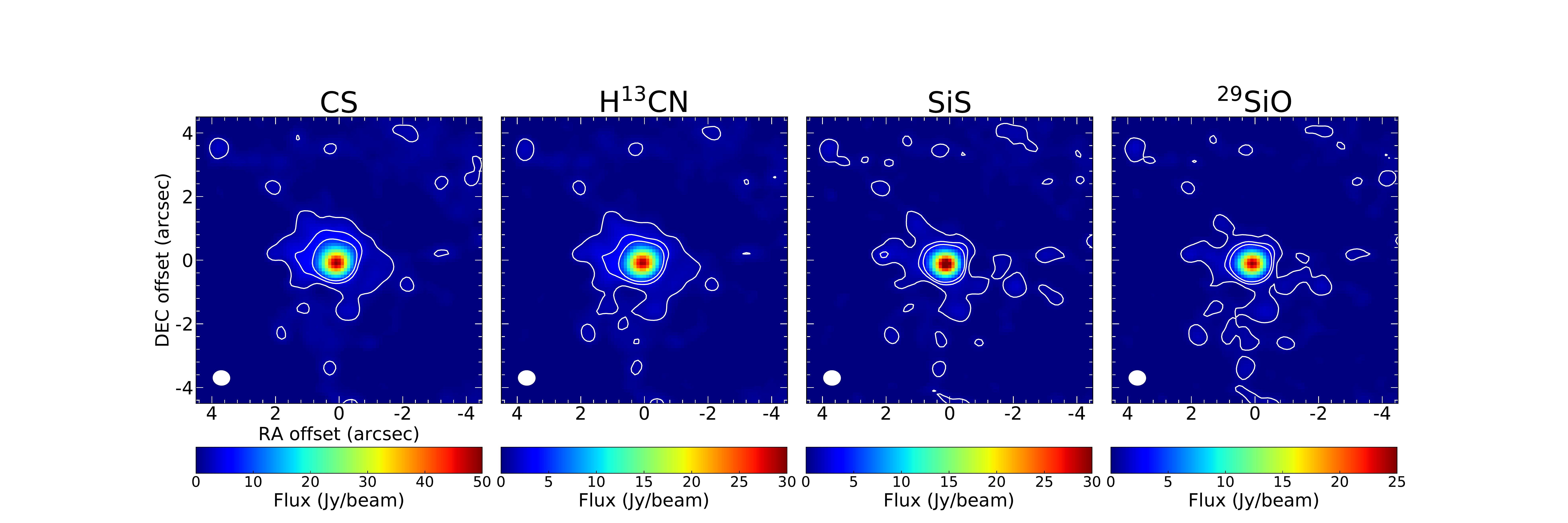}
   \caption{ALMA integrated intensity (moment 0) maps of CS\,(7--6), H$^{13}$CN\,(4--3), SiS\,(19--18), and $^{29}$SiO\,(8--7)  emission lines around W\,Aql. The ALMA beam (0.55$^{\prime\prime\,} \times$ 0.48$^{\prime\prime\,}$, PA 86.35$^\circ$) is given in the lower-left corner. Contours are marked in white for 1, 3 and 5$\sigma$ rms. North is up and east is left. }
              \label{fig:strongmoments}%
    \end{figure*}
    
           \begin{figure*}
   \centering
      \includegraphics[trim={7.5cm 0cm 7.5cm 3cm},clip,width=0.98\textwidth]{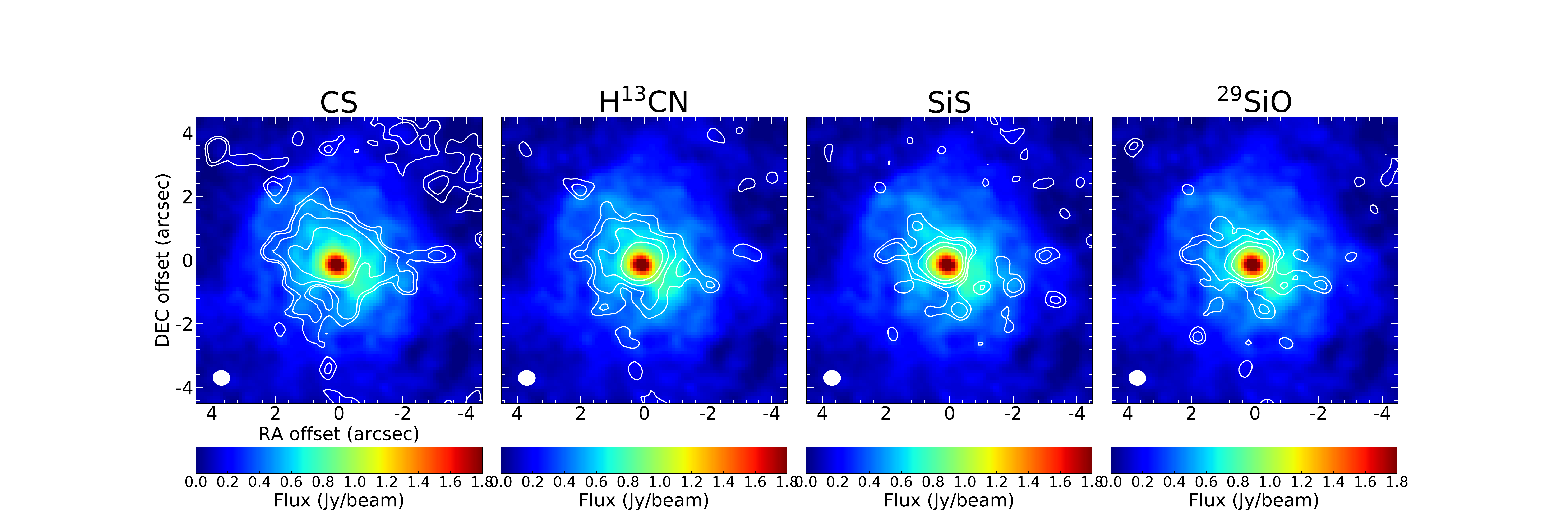}
   \caption{ALMA CO image (color) overlaid with contours (white) of the CS\,(7--6), H$^{13}$CN\,(4--3), SiS\,(19--18), and $^{29}$SiO\,(8--7) emission, shown at stellar velocity (about -21\,km~s$^{-1}$). The ALMA beam for the CS, H$^{13}$CN, $^{29}$SiO and SiS lines (0.55$^{\prime\prime\,} \times$ 0.48$^{\prime\prime\,}$, PA 86.35$^\circ$) is given in the lower-left corner. Contours are plotted for 3, 5, 10 and 20$\sigma$ rms. North is up and east is left. }
              \label{fig:cocomp}
    \end{figure*}

\subsection{Morphology of spectral line emission}

For the strong spectral line emission -- shown in moment-zero maps in Fig.\,\ref{fig:strongmoments} and channel maps in Appendix\,\ref{sec:appendix1} -- the peak emission seems to be roughly spherical, concentrated at the stellar position, and is spatially resolved. Additionally, we see a fainter and slightly elongated circumstellar emission component extending out to roughly 2 -- 3$^{\prime\prime\,}$. The elongation is oriented in the north-east and south-west (NE/SW) directions, which is also the direction of the binary orbit. We compare the molecular emission of the four strong spectral lines with the CO\,(3--2) emission reported by \citep{Ramstedt2017} in Fig.\,\ref{fig:cocomp} and find that the emission features along the NE/SW elongation are equally represented in the CO emission as well, with emission peaks in the molecular lines coinciding with CO emission peaks or clumps.

\subsection{Spectral features}

\begin{figure}[t]
   \centering
   \includegraphics[trim={0.3cm 0.5cm 0.3cm 0.2cm}, width=0.49\textwidth]{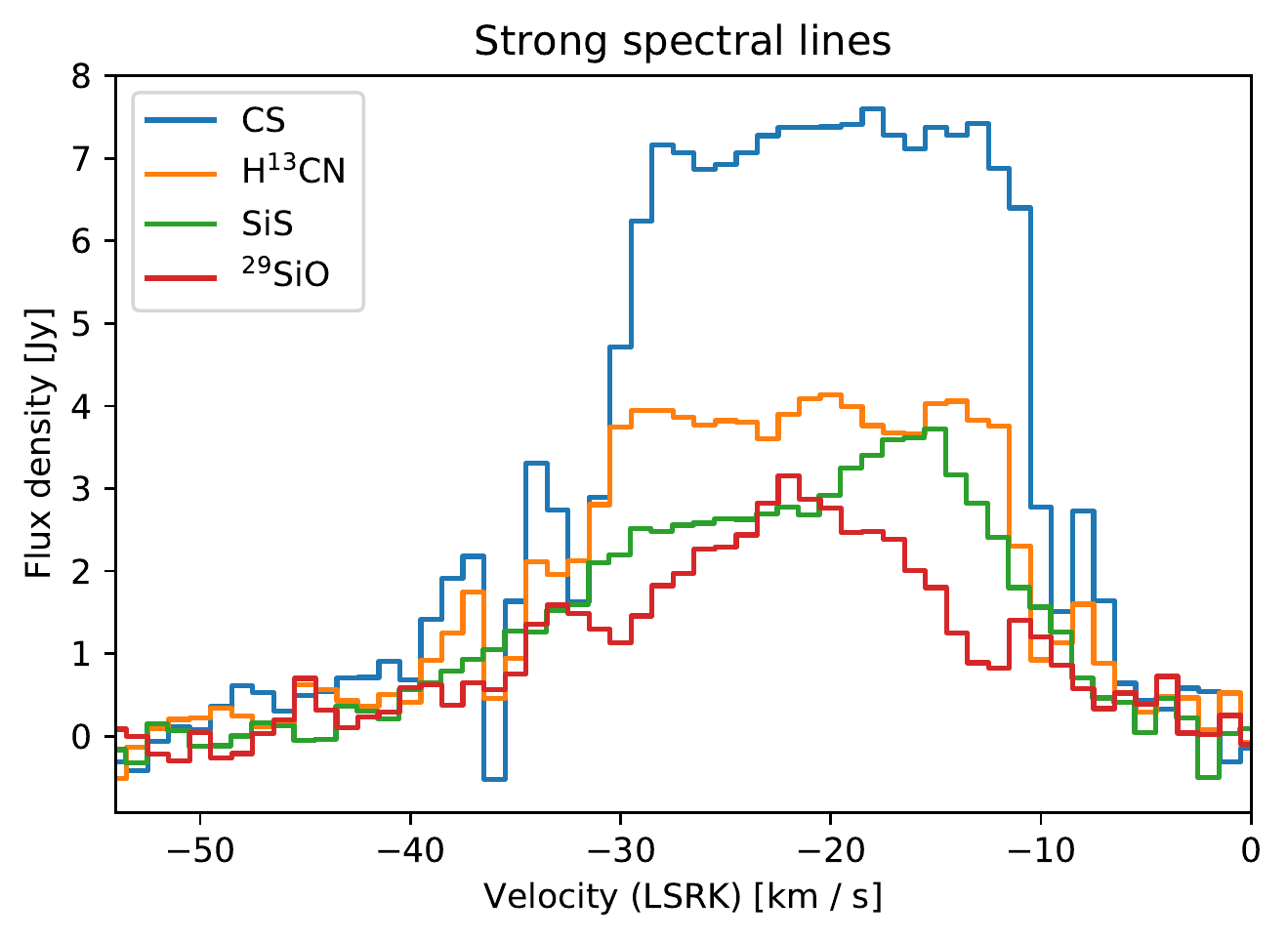} 
   \caption{Spectra of the four strongest spectral lines -- CS, H$^{13}$CN, SiS and $^{29}$SiO --  measured within a circular 5$^{\prime\prime\,}$ aperture centered on the star. The stellar velocity is at about -21\,km~s$^{-1}$.}
   \label{fig:strongspec}
\end{figure}

The four strongest emission lines (Fig.\,\ref{fig:strongspec}) all show a slight but noticeable excess in blue-shifted emission between approximately -35 and -40\,km/s. This cannot be seen in the channel maps (Appendix\,\ref{sec:appendix1}). Additionally, the SiS line shows a strong asymmetry with an excess in red shifted emission relative to the stellar rest frame at around -21\,km~s$^{-1}$. This feature can possibly be seen also in the channel maps (Fig.\,\ref{fig:SiSmap}), where between roughly -20 and -14\,km~s$^{-1}$~an elongation of the bright peak emission to the east can be seen. It is most prominent at about -16\,km~s$^{-1}$, which coincides with the peak of the asymmetry in the spectrum. Furthermore, the $^{29}$SiO profile shows two secondary features at about -32\,km~s$^{-1}$ and -11\,km~s$^{-1}$.

Previous models on multiple spectral lines by \citet{Danilovich2014} report a stellar velocity (extracted from the line center velocity), $v_{lsr}$,  of -23\,km/s, which is backed up by and consistent with several other studies, reporting a very similar velocity \citep[e.g.,][]{Bieging1998,Ramstedt2009,De-Beck2010,Mayer2013,De-Nutte2017}. We note that, investigated separately from previously detected lines, the reported lines observed with ALMA seem to be generally redshifted to this stellar velocity, and a $v_{lsr}$ of -21\,km/s \citep[such as that calculated by][]{Mayer2013} is more fitting. Taking the ALMA line peak velocities of all observed lines and calculating an average line peak velocity, we also arrive at a $v_{lsr}$ of -21\,km/s, but the asymmetry of the lines and deviation from parabolic line profiles is very line-specific (as seen in Fig.\,\ref{fig:strongspec}).

\subsection{uv fitting}
\label{sec:uvfitting}

Since the imaging process of interferometric data involves initial knowledge of the source geometry and image artefacts are inevitably introduced during deconvolution, it is reasonable to analyze the observations already in the uv plane, directly working with the calibrated visibility data. For this reason, we use the CASA tool \textit{uvmultifit} \citep{Marti-Vidal2014}, to fit geometric flux models to the bright emission regions of all spectral lines. Since the TP data combination for this project is done in the image plane and not in the uv plane, we can only fit to the combined 12\,m main array and ACA data. This is a reasonable approach, since the maximum recoverable scale for the used 12\,m main array and ACA configuration is 19$^{\prime\prime\,}$, and we concentrate our analysis on the central few arcseconds.

\begin{figure*}[htbp] 
   \centering
   \includegraphics[trim={7.0cm 5cm 7.5cm 3cm},clip,width=0.98\textwidth]{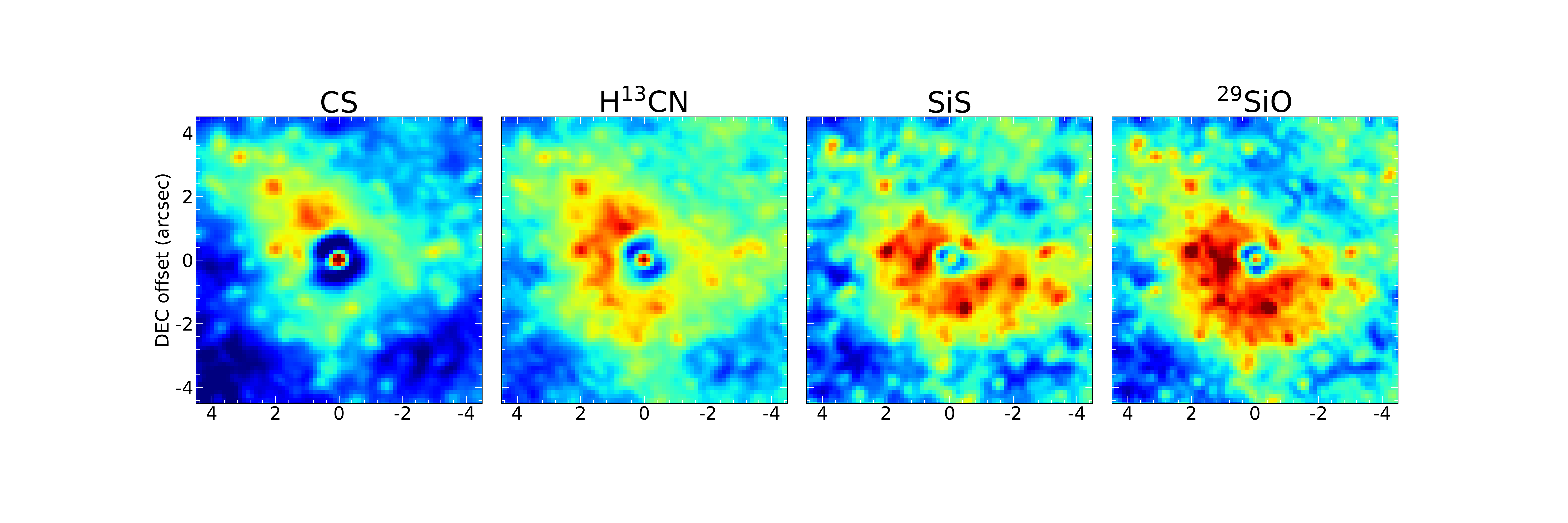} 
      \includegraphics[trim={7.0cm 0cm 7.5cm 5.5cm},clip,width=0.98\textwidth]{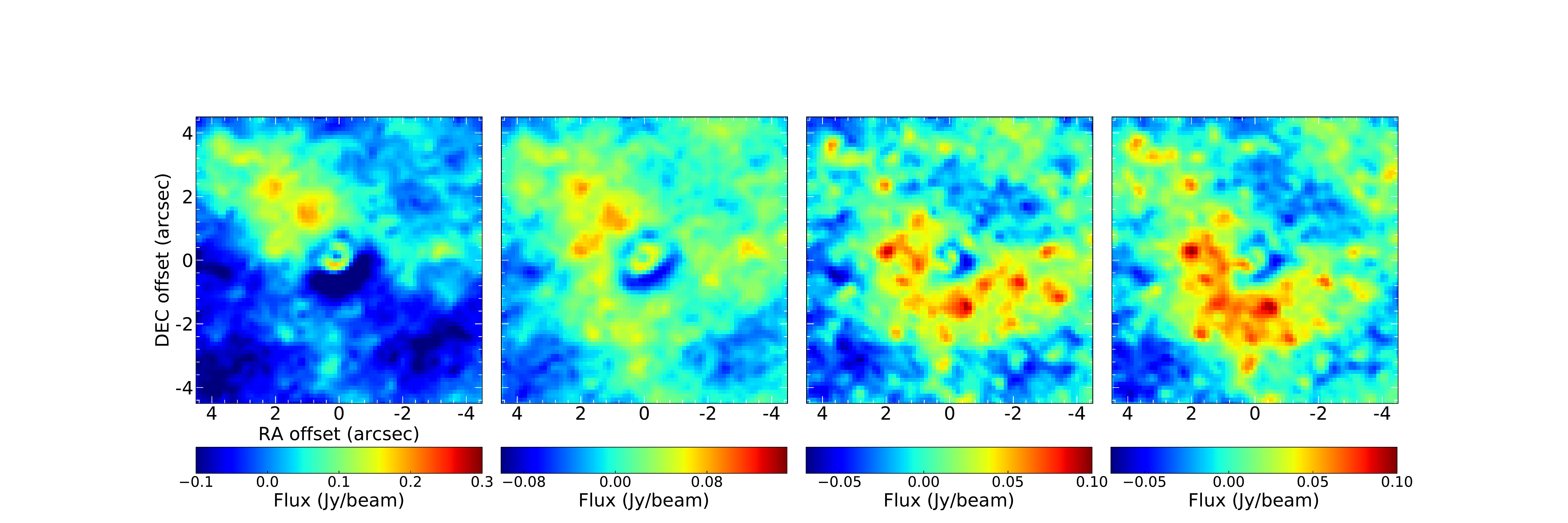} 
   \caption{Imaged residuals of the four strong emission lines after subtraction of the model visibilities of a uv fit. Only the central channel at -21\,km/s is shown. Top row: Residuals after the subtraction of the uv fit of a circular Gaussian; Bottom row: Residuals after the subtraction of the uv fit of a circular Gaussian and a delta function.}
   \label{fig:image-residuals}
\end{figure*}

\begin{figure*}[htbp] 
   \centering
   \includegraphics[width=1.0\textwidth]{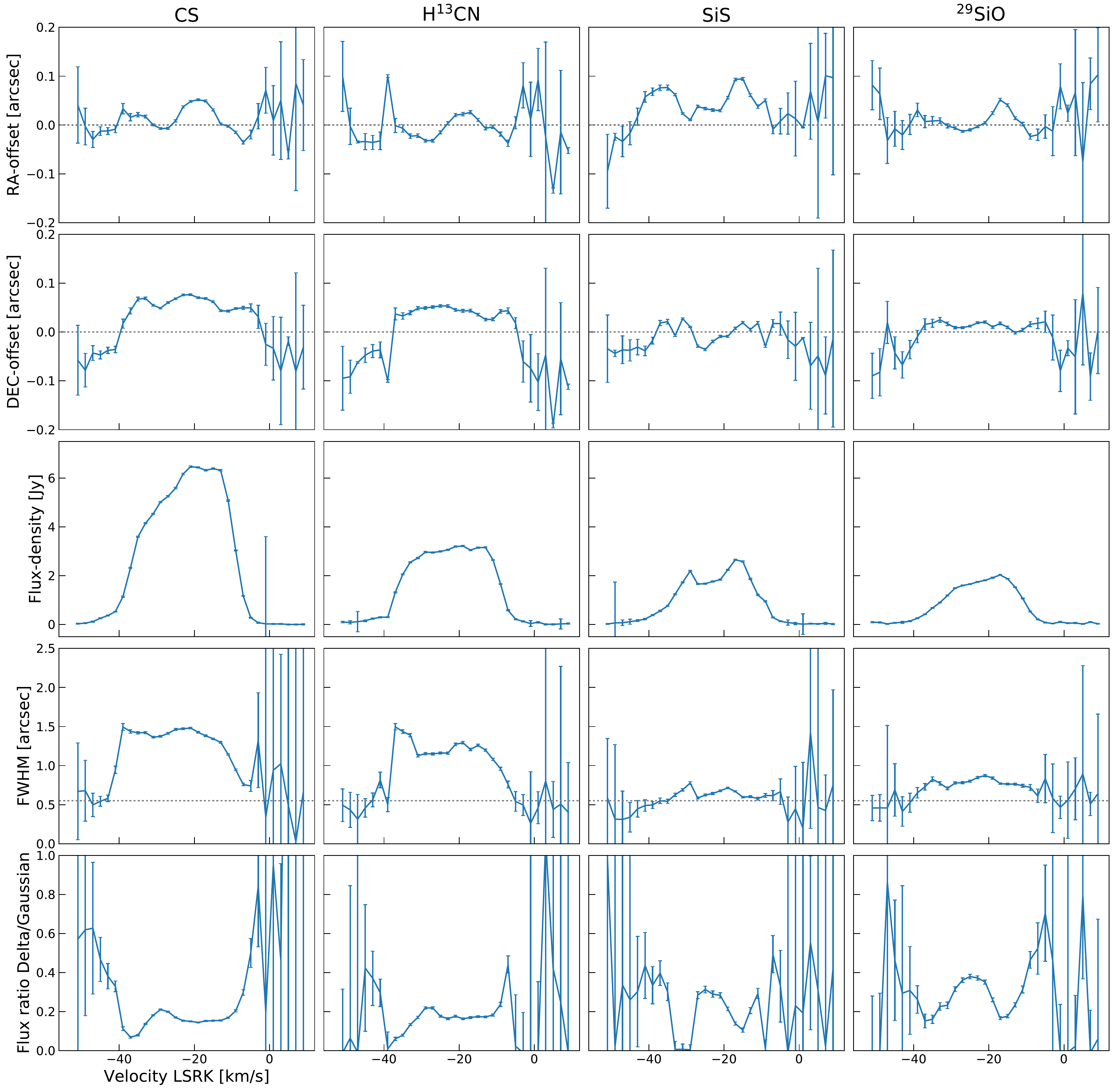} 
   \caption{uv-fitting of the four strong spectral lines detected in the inner CSE of W\,Aql. First row: RA offset position of the fitted Gaussian. Second row: DEC offset position. Third row: Fitted flux-density of circular Gaussian for each velocity bin of 2 km/s. Fourth row: Size of best-fitting circular Gaussian (FWHM). Fifth row: Flux ratio of fitted delta function to Gaussian. For reference, the effective beam size in the observed images of 0.55$^{\prime\prime\,} \times$ 0.48$^{\prime\prime\,}$ is indicated by a gray dashed line in the fourth row. The gray dashed lines in the two top rows denote the zero point of position offset in RA and DEC, respectively.}
   \label{fig:uvfitting-stronglines}
\end{figure*}

Therefore we conclude that for the analysis of the compact, bright emission it is sufficient to use only the main array and ACA data. As a first step, we fit a compact, circular Gaussian component to the data, which is described by its position (relative to the current stellar position of RA\,19:15:23.379 and DEC\,–07:02:50.38 J2000.0 \citep{Ramstedt2017}), flux density, and FWHM size. After subtracting the model visibilities from the observed ones, we imaged the residuals to verify the goodness of the Gaussian fit. The resulting residual images for the four strong spectral lines are shown in the top row of Fig.\,\ref{fig:image-residuals}.  Apart from relatively faint, extended emission surrounding the subtracted compact Gaussian, which can be seen for all four spectral lines, a very compact emission peak remains at the stellar position, which can best be seen in the residual CS and H$^{13}$CN emission. We succeed in fitting the compact, central emission by adding a delta function to the circular Gaussian, and arrive at significantly better residual images (see Fig.\,\ref{fig:image-residuals}, bottom). Figure\,\ref{fig:uvfitting-stronglines} shows the uv-fitting parameter results for the Gaussian position offset, the Gaussian flux density, and Gaussian FWHM fit parameters, as well as the flux ratio between the added delta component and the Gaussian, for the four strong spectral lines.

We note, however, that the true intensity distribution of a uniformly expanding envelope is not strictly Gaussian. When the emission is optically thin, the intensity distribution $\propto r^{-1}$, which cannot be straightforwardly implemented for use in a uv analysis. When the emission becomes optically thick, the inner part of the distribution flattens, improving a Gaussian approximation. The distribution in the outer envelope can be adequately described using a Gaussian distribution and the deviation mainly becomes apparent in the inner region, close to the resolution limit of our observations. The deviations in this part explain the better fits obtained when including a delta-function, and the strength of the delta function depends on the line optical depth.

The offset of the fitted Gaussian position to the stellar position is well below the spatial resolution of the images and a clear general trend of the offsets for all four lines is barely visible. We note, however, that the photocenter seems to shift with increasing flux density towards slightly larger RA and DEC.

For all four lines, the FWHM size for channels with velocities larger than 0\,\,km~s$^{-1}$~shows very large error bars, meaning that the fit does not converge in those channels and the estimated size of the emission region is very uncertain. The flux density fits show very similar line profiles to the observed ones (a direct comparison is given in Sect.\,\ref{sec:uvspeccomp}), with the exception of the $^{29}$SiO line profile, which does not show the additional peaks at blue- and red-shifted velocities in the Gaussian fit. 
The CS, H$^{13}$CN and $^{29}$SiO lines each show a peak in the emission size at blue velocities, which cannot be directly associated with any features in the flux density profile. The only similarity to note here is that the size of the emission region changes at the velocities that belong to the slightly asymmetric and weaker part of the flux density profile. On the other hand, the asymmetric feature in the SiS line profile is associated with a slight increase in emission size. A very similar asymmetry in the SiS line has been observed for the carbon star RW~LMi \citep{Lindqvist2000}.

We note that the fitted FWHM of the SiS line lies very close to the imaging resolution at some velocities, meaning that the uv fit is close to over-resolving the emission. We refer to Appendix\,\ref{appendixuv} for a discussion about super-resolution in uv fitting, as well as the uncertainties attached to the fits.

As expected, the uv-fitting of the four weaker spectral lines is more uncertain and especially for the extreme velocity channels the emission size cannot be constrained confidently. The SiO v=1 and $^{30}$SiS lines seem to be slightly resolved, while the SiO v=2 and SiS v=1 lines are unresolved in the image plane, and the signal-to-noise ratio (S/N) for all weak lines is too low for confident fitting. Therefore we refrain from presenting inaccurate uv-fitting results for the weak spectral lines.

\subsection{Comparison between uv-fitted and observed spectral shape}
\label{sec:uvspeccomp}

\begin{figure*}[htbp] 
   \centering
   \includegraphics[width=1.0\textwidth]{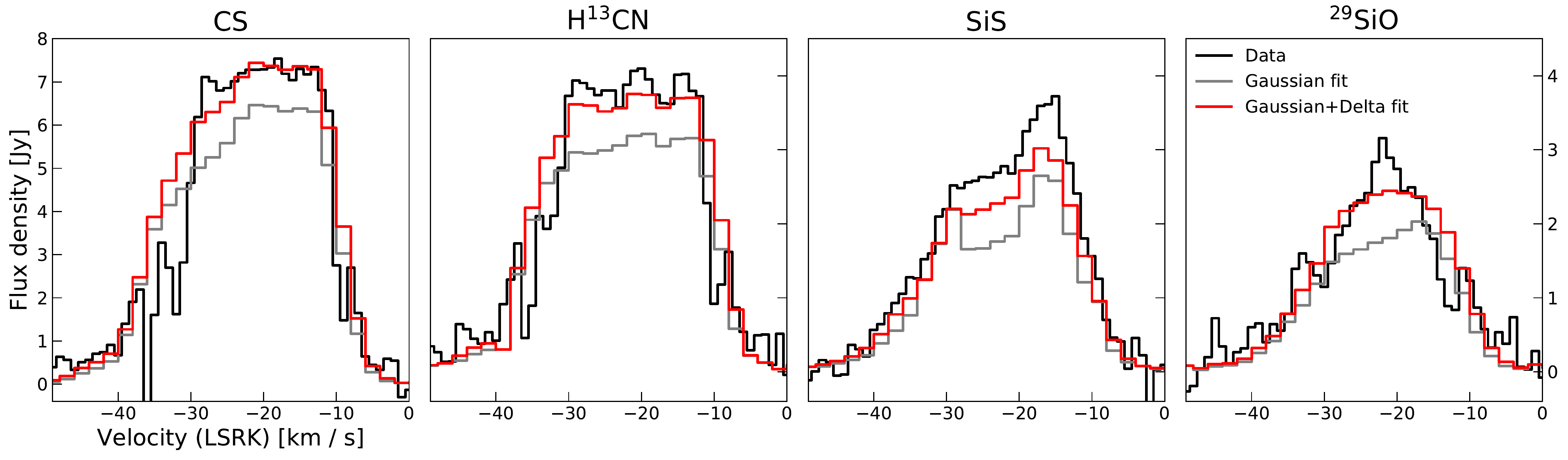} 
   \caption{Comparison of the observed spectra (black) with the uv-fitted spectra of the Gaussian fit (gray) and the fit including a Gaussian plus a delta function (red). The observed spectra were extracted from the images with an aperture of 5$^{\prime\prime\,}$ diameter. The effective beam size in the observed images is 0.55$^{\prime\prime\,} \times$ 0.48$^{\prime\prime\,}$. We note that the y-axis scales differently for the first spectral line, which is the strongest. The respective scales are given on the left-hand side for the first line, and on the right-hand side for the remaining lines. }
   \label{fig:spec-comp-strong}
\end{figure*}

We compare the uv-fitted spectra with the observed spectra extracted from the beam convolved images with an aperture of 5$^{\prime\prime\,}$ diameter  to compare the spectral contribution of the compact Gaussian and delta function emission components with the fainter, extended asymmetric emission component seen in the residual images (Fig.\,\ref{fig:image-residuals}). Figure \ref{fig:spec-comp-strong} shows the compared spectra for the strong lines.  To illustrate the
flux contribution of the compact delta component, we separately show the spectrum of a uv fit with just a circular Gaussian and a spectrum of the uv fit with a circular Gaussian and added delta function component.

For all four lines, we see that the fraction of recovered flux is significantly lower when no delta component is added to the uv fit. The model including a Gaussian and a delta component recovers almost the total flux for the CS and H$^{13}$CN lines, and the majority of flux for the SiS and $^{29}$SiO lines. While for CS and H$^{13}$CN the contribution of the delta component seems to be roughly of the same strength for all velocities, the contribution of the delta component compared to the Gaussian is slightly higher at blue shifted velocities for SiS and $^{29}$SiO.

We conclude that the overall emission line profile can be recovered well by uv models of a Gaussian and a delta component, while the remaining small deviations can be attributed to the residual extended and asymmetric emission.

\section{Radiative transfer modeling}
\label{sec:radtransfer}

\subsection{Modeling procedure}

To model the molecular emission, we used ALI, an accelerated lambda iteration code \citep[first introduced, and later applied and improved by ][ respectively]{Rybicki1991,Maercker2008}, following the same procedure as described in \citet{Danilovich2014}. The stellar parameters used for the models are given in Table\,\ref{tab:stellarparameters}. To constrain our models, we used the azimuthally averaged radial profile from the ALMA observations (convolved with a 0.6$^{\prime\prime\,}${} beam) and the ALMA line profiles convolved with a 4$^{\prime\prime\,}${} beam, and additional archival single-dish data listed in Table\,\ref{archivalobs}. We note that the models are primarily fitted to the ALMA data, with the integrated flux of the archival single-dish data being used -- in most cases --  to discriminate between two equally good fits to the ALMA data. Our model uses the CO modeling results obtained by \citet{Danilovich2014} and refined by \citet{Ramstedt2017}. For a distance of 395\,pc, 1$^{\prime\prime\,}${} corresponds to 5.91$\times 10^{15}$\,cm or $1\times 10^{15}$\,cm corresponds to 0.17$^{\prime\prime\,}${}.

In all cases, we assumed a Gaussian profile for the molecular fractional abundance distribution, centred on the star, following
\begin{equation}
f(r) = f_0 \exp\left( -\left( \frac{r}{R_e} \right)^2\right)
,\end{equation}
where $f_0$ is the abundance relative to H$_{2}$ at the inner radius of the molecular CSE and $R_e$ is the $e$-folding radius, the radius at which the abundance has dropped off by $1/e$.

The Gaussian models have two free parameters per molecular species: $f_0$, the central fractional abundance, and $R_e$, the $e$-folding radius. When considering the ALMA radial profiles, we constrain our models using the central position and offset positions with spacings of 0.1$^{\prime\prime\,}$ starting at 0.05$^{\prime\prime\,}$ and extending to 0.75$^{\prime\prime\,}$, although we include plots showing the models and ALMA profiles out to 2.55$^{\prime\prime\,}$. The error bars on the ALMA observations are derived from an azimuthal average over the observed radial profile.
Regarding optical depth, we note that $^{30}$SiS and $^{29}$SiO are optically thin, CS and $^{28}$SiS are mostly optically thin, with the exception of the very innermost regions of CSEs, $^{28}$SiO is mostly optically thick, H$^{12}$CN is very optically thick, and H$^{13}$CN is optically thick in the shell, but optically thin elsewhere.

\begin{table}
\caption{Stellar parameters used for radiative transfer modeling.}
\label{tab:stellarparameters}

\begin{center}
\begin{tabular}{llrcc}

\toprule
\multicolumn{2}{c}{Model parameters}  \\
\multicolumn{2}{c}{W\,Aquilae}  \\
\midrule
Distance & 395\,pc\,\tablefootmark{a} \\
Effective temperature & 2300\,K  \\
Gas mass-loss rate & $3\times 10^{-6}$ \,M$_{\odot}$/yr\,\tablefootmark{b}\\
Dust/gas mass ratio & $2\times 10^{-3}$  \\
Luminosity & 7500\,L$_{\odot}$\,\tablefootmark{c} \\
Stellar mass & 1\,M$_{\odot}$ \\
Gas expansion velocity & 16.5\,km/s\,\tablefootmark{d} \\
Stellar velocity (LSRK) & -21\,km/s\,\tablefootmark{d}\\

\bottomrule
\end{tabular}
\end{center}
\tablefoot{All parameters are adopted from \citet{Danilovich2014}.\\
\tablefoottext{a}{Derived from period-magnitude relation \citep{Whitelock2008} and 2MASS $K$ band magnitude \citep{Cutri2003}.}
\tablefoottext{b}{\citet{Ramstedt2017}}
\tablefoottext{c}{Derived from period-luminosity relation \citep{Glass1981}.}
\tablefoottext{d}{Measured from CO line data \citep{Danilovich2014}.}
}

\end{table}

\begin{table}[tp]
\caption{Results of radiative transfer modeling.}
\label{modellingresults}
\begin{center}
\begin{tabular}{lccccc}
\toprule
Molecule                & $f_0$         & \multicolumn{1}{c}{{\parbox{1cm}{$R_e$ \\ (cm)}}} &  \multicolumn{1}{c}{{\parbox{1.5cm}{Overdensity \\abundance}}}\\
\midrule
CS & $(1.20 \pm 0.05) \times 10^{-6}$ & $7.0\times 10^{15}$& $\times2$\\
SiS & $(1.50 \pm 0.05)\times 10^{-6}$ & $6.0\times 10^{15}$&$\times3$\\
$^{30}$SiS & $(1.15 \pm 0.05)\times 10^{-7}$ & $3.5\times 10^{15}$&$\times1$\\
$^{29}$SiO & $(2.50 \pm 0.10)\times 10^{-7}$ & $1.4\times 10^{16}$&$\times1$\\
H$^{13}$CN& $(1.75 \pm 0.05)\times 10^{-7}$ & $4.7\times 10^{16}$ &$\times1$\\
\bottomrule
\end{tabular}
\end{center}
\tablefoot{$f_0$ is the central abundance relative to H$_{2}$ and $R_e$ is the $e$-folding radius. The last column indicates if and with what factor the abundance in the overdensity region has to be increased by multiplication for achieving the best fit.}
\end{table}%

%

\subsection{Modeling results}
\label{sec:results}

When modeling the ALMA radial profiles, we found that for most molecules there was a ``tail'' in the radial profile, from around 1$^{\prime\prime\,}$ outwards, which could not be fit with a Gaussian abundance profile in a smoothly accelerating CSE model. We first attempted to find a good fit with a non-Gaussian abundance profile, but such models that fit the observations well were not physically realistic, with increasing abundances with radius. Instead, we found that adjusting the wind density over a small region, as could be caused by a binary induced, spiral-like overdensity, for example, greatly improved our model fits to the entire profile. For a smoothly accelerating wind with no overdensities, the H$_2$ number density at radius $r$ is given by
\begin{equation}\label{nh2}
n_{\mathrm{H}_2} (r) = \frac{\dot{M}}{4\pi r^2 m_{\mathrm{H}_2} \upsilon(r)}
,\end{equation}
where $\dot{M}$ is the mass-loss rate, $m_{\mathrm{H}_2}$ is the mass of H$_2$ and $\upsilon(r)$ is the radial velocity profile. To adjust the H$_2$ number density, we multiplied a section of the smooth-wind radial profile (given by the mass-loss rate determined by \citet{Ramstedt2017}, listed in Table \ref{tab:stellarparameters}, and the velocity profile defined by \citet{Danilovich2014}) by integer factors up to 10. The result that did not over-predict the ``tail'' for any molecule was chosen as the best model. We found that the inner regions of the radial intensity profiles were not significantly affected by the inclusion of this overdensity and the central abundances and $e$-folding radii did not need to be adjusted when adding the overdensity. 

\begin{figure}[t]
   \centering
   \includegraphics[width=0.49\textwidth]{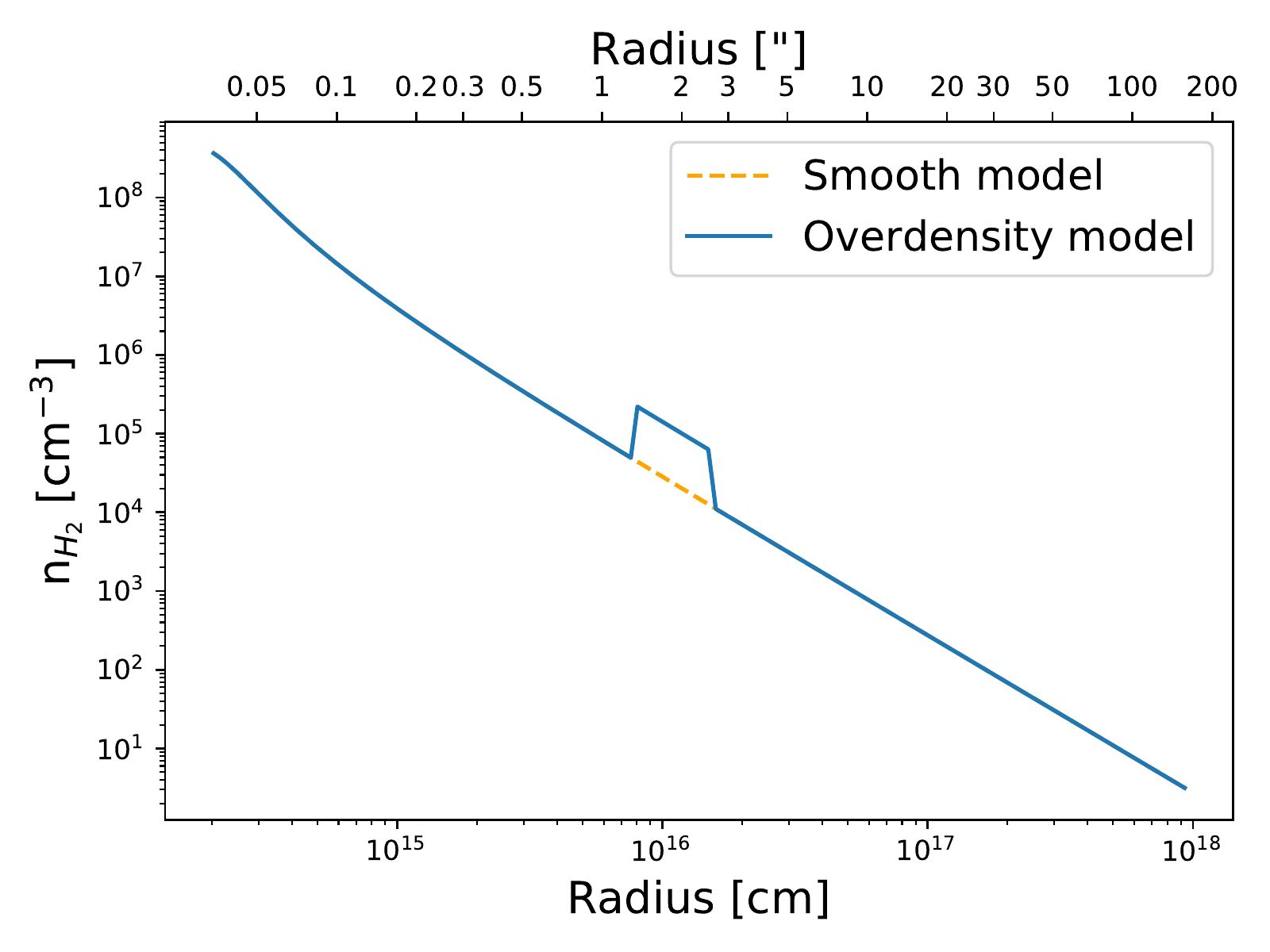}
      \caption{Radial $\mathrm{H}_2$ number density for both the smooth wind model and the overdensity model.}
         \label{fig:nh2plot}
   \end{figure}

The best fit for all molecules was found for an overdensity of a factor of five between $8\times 10^{15}$~cm and $1.5\times 10^{16}$~cm.
This reproduces the observed radial profiles for H$^{13}$CN and $^{29}$SiO well, without adjustments to the abundance profiles determined from the smooth-wind models. For SiS and CS, we find that the overdensity model improves the fit to the tail, but not sufficiently to reproduce it entirely. The $\mathrm{H}_2$ number densities for both the smooth wind and overdensity models are plotted in Fig.\,\ref{fig:nh2plot}.
While testing our models, we found that an overdensity factor of ten for SiS or eight for CS were better fits, but these overdensity models strongly over-predicted the H$^{13}$CN and $^{29}$SiO ``tails''. Instead, to find the best fits for SiS and CS, we additionally altered their abundances in the overdense region. The scaling factors of the altered abundances are listed in the last column of Table\,\ref{modellingresults}. We find this approach justified because the same wind overdensity must be applied to all modeled molecules.

   \begin{figure}[t]
   \centering
   \includegraphics[width=0.48\textwidth]{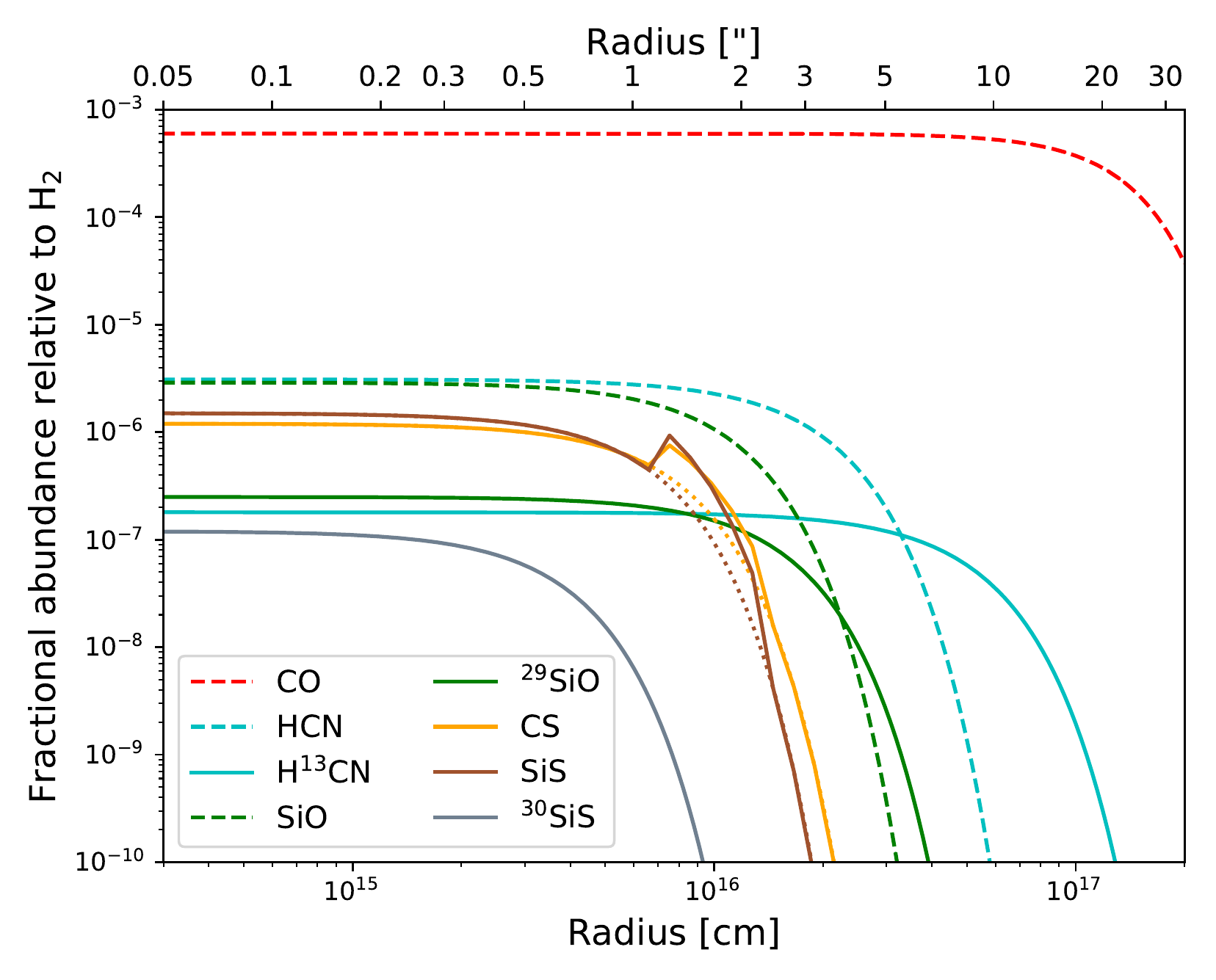}
      \caption{Radial abundance profile. \textit{Full lines:} molecules observed and modeled in this study, using the overdensity model. Where the abundances differ from the best fitting smooth model, the smooth model is indicated by a dotted line of the same color. \textit{Dashed lines:} molecules modeled by \citet{Danilovich2014}.}
         \label{fig:abundanceprofile}
   \end{figure}

In Table\,\ref{modellingresults}, we list the best fit models for both the smooth wind model and the overdensity model, based on the vibrational ground state observations of the respective molecules. The presence of the tail and our overdensity model solution complicates the error analysis of the envelope size. However, if we exclude the overdensity model and discuss only the smooth model, we find that the constraints on the envelope size are tighter for the molecules with less prominent tails and less so for the molecules with the most pronounced tails, such as H13CN. In fact, we found that it was not possible to fit the tail of H13CN simply by increasing the Gaussian e-folding radius, while also providing a good fit to the inner radial points. A Gaussian with a comparable size to the CO envelope (unrealistic in the case of H13CN) would still not fit the tail. Hence the need for the overdensity model.

\begin{figure*}[htbp] 
\centering
   \includegraphics[ width=0.49\textwidth]{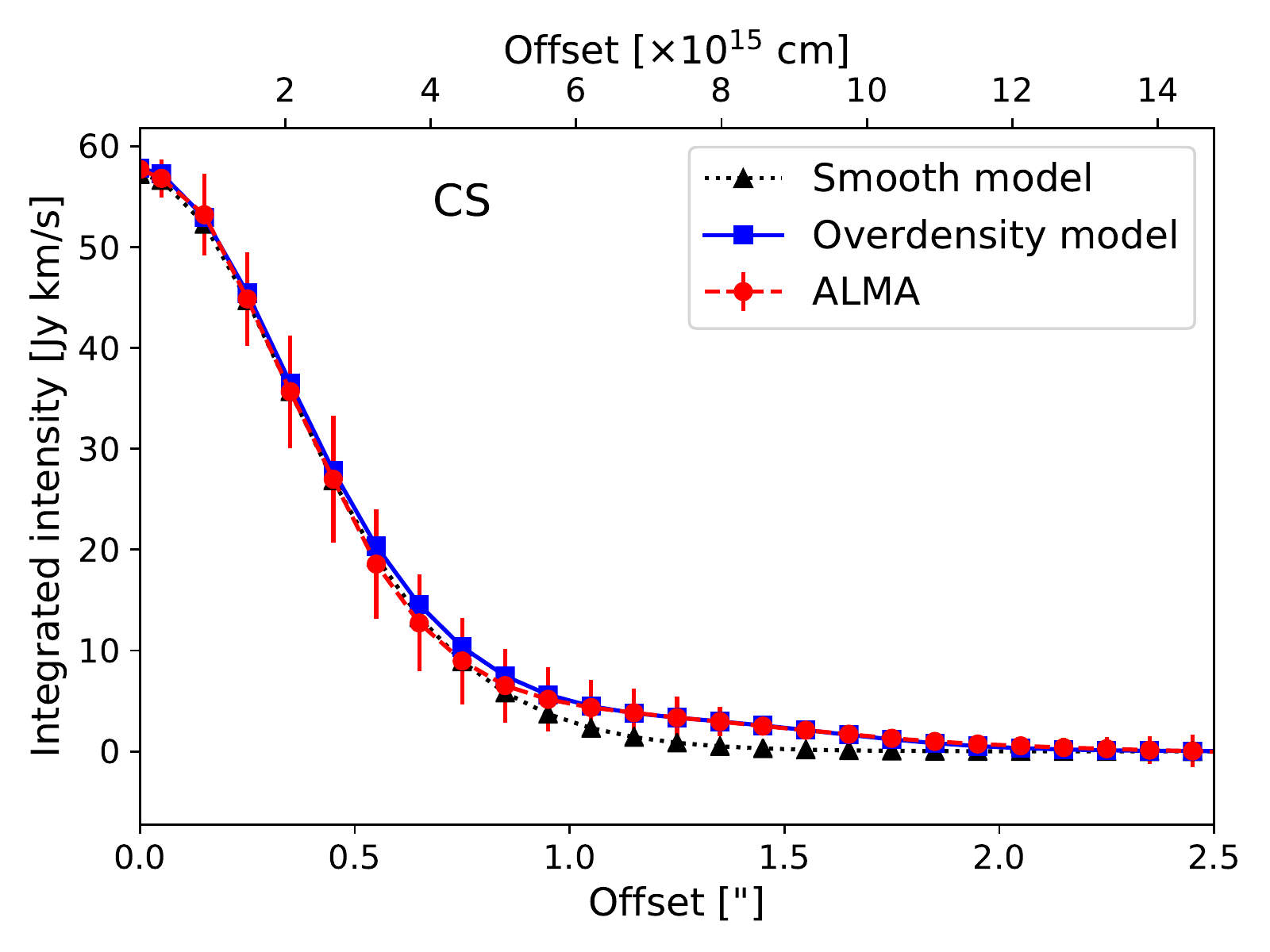}
   \includegraphics[ width=0.49\textwidth]{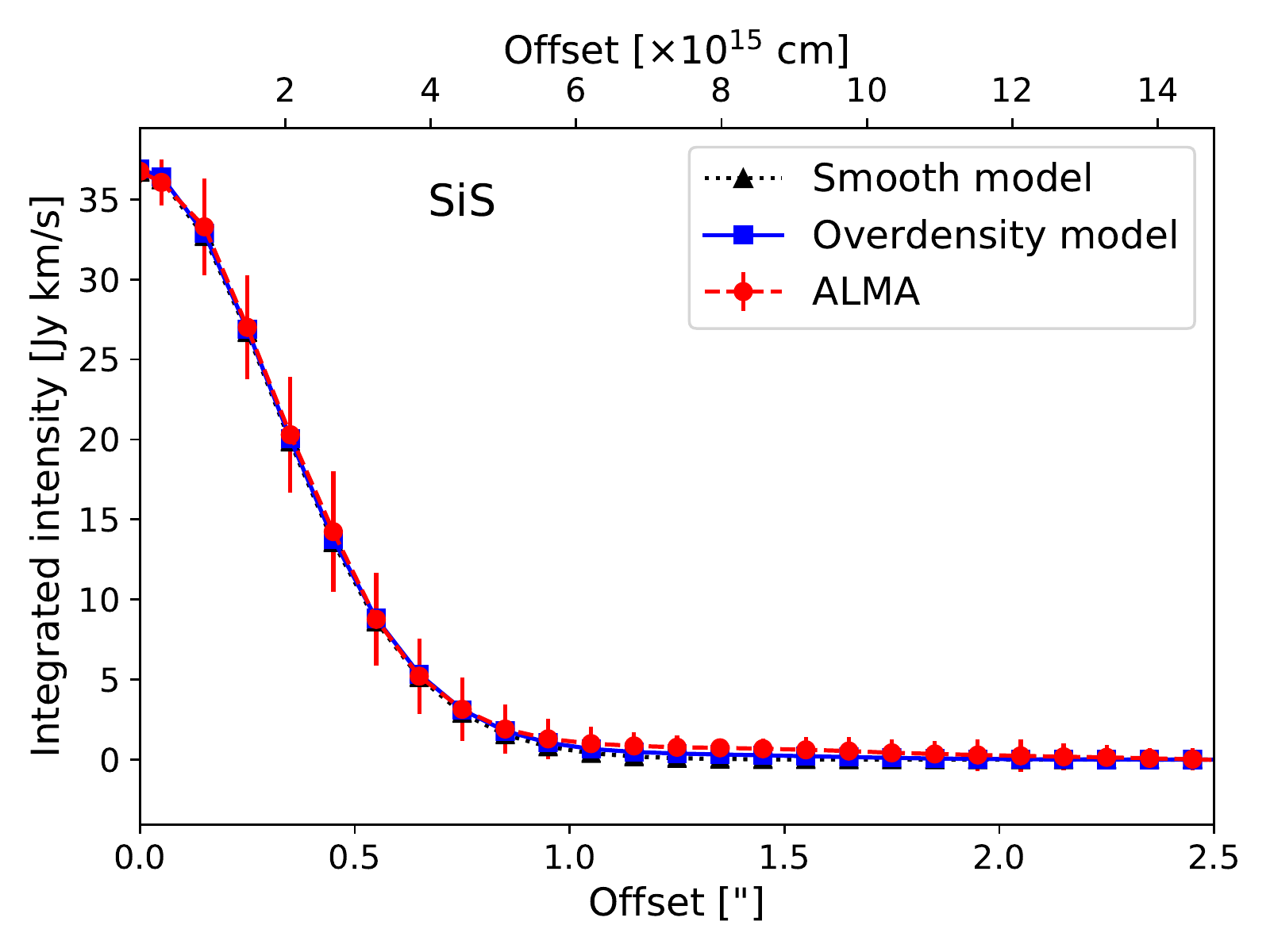}
   \includegraphics[ width=0.49\textwidth]{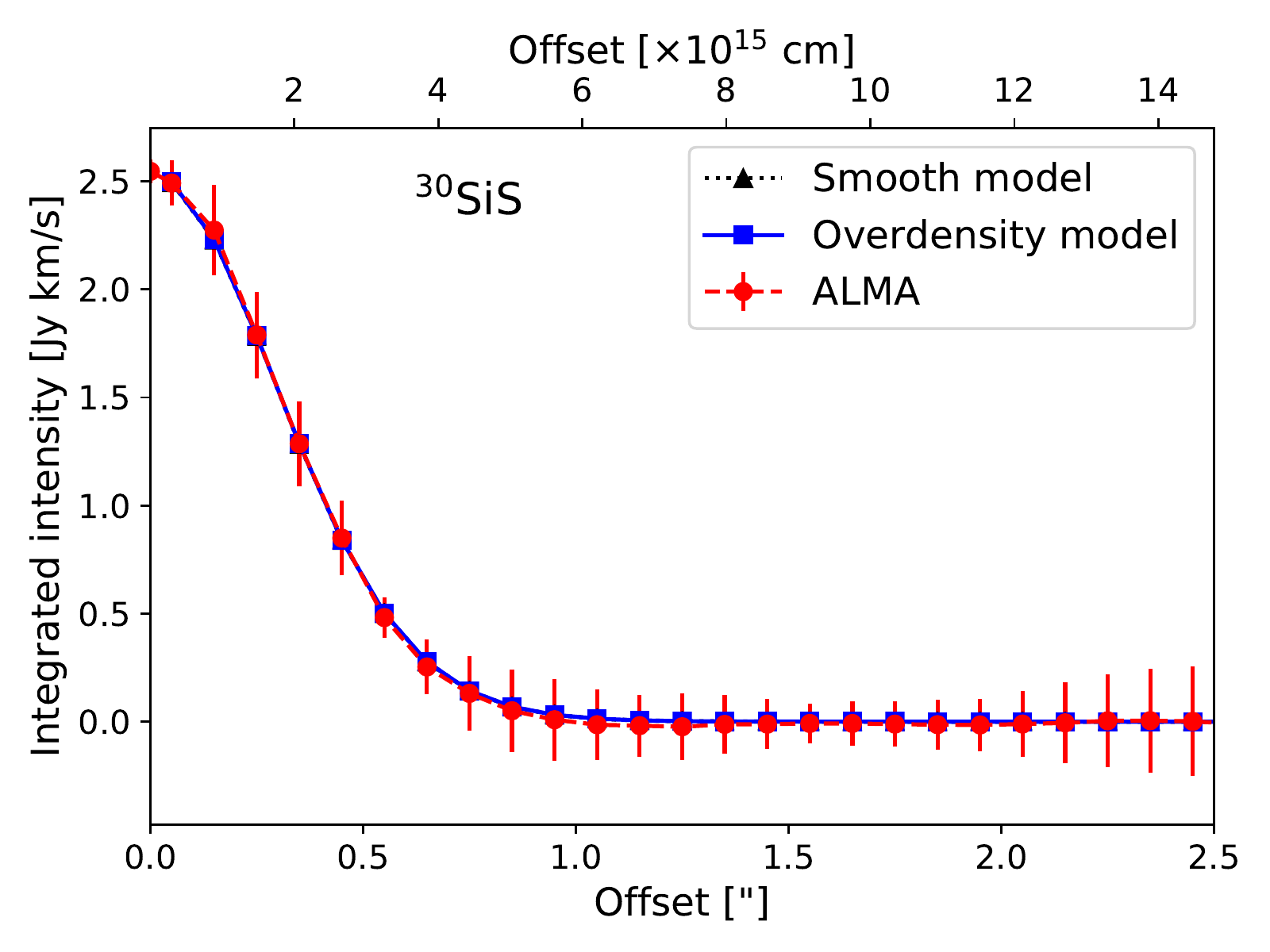}
   \includegraphics[ width=0.49\textwidth]{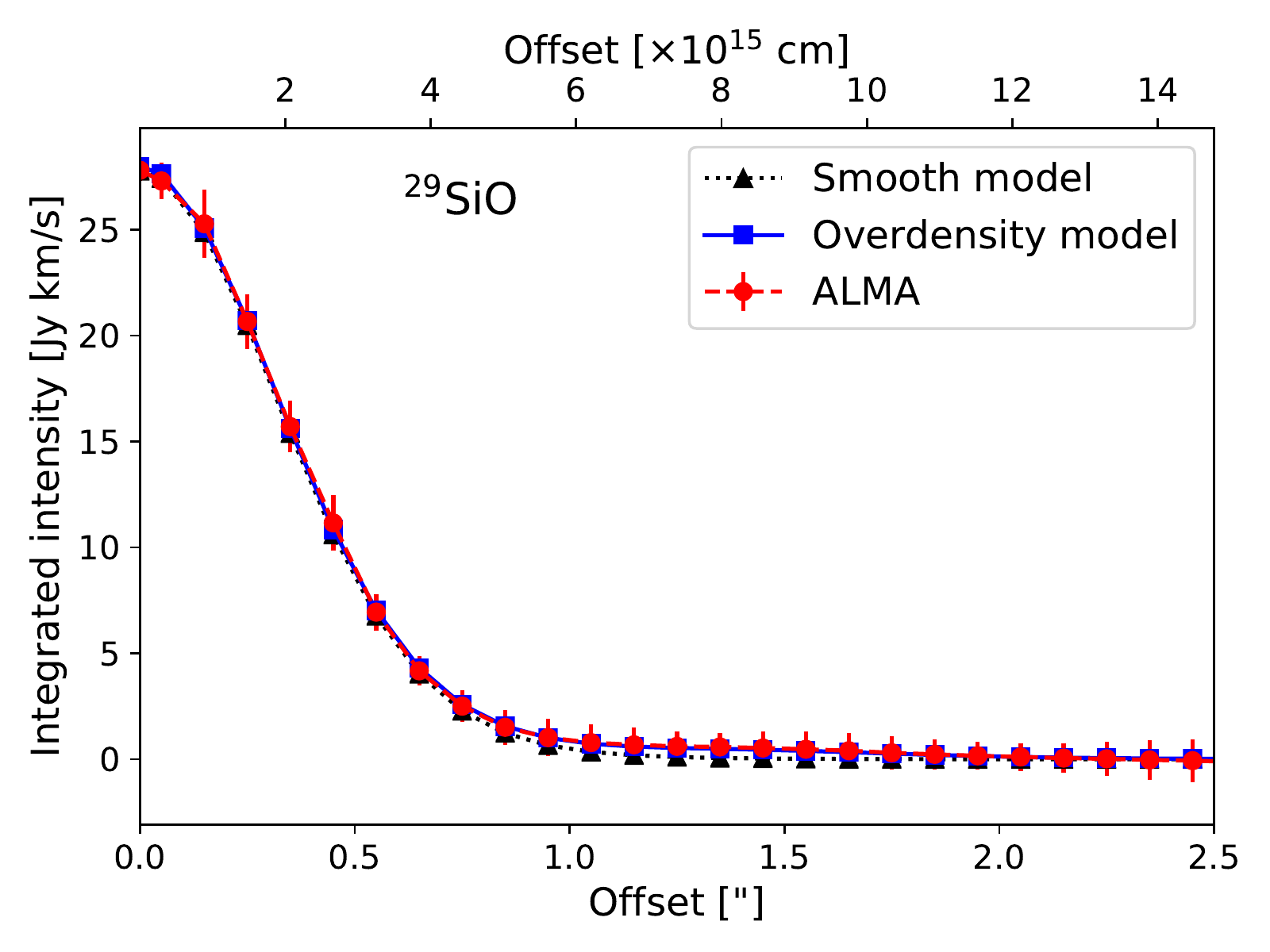}
   \includegraphics[ width=0.49\textwidth]{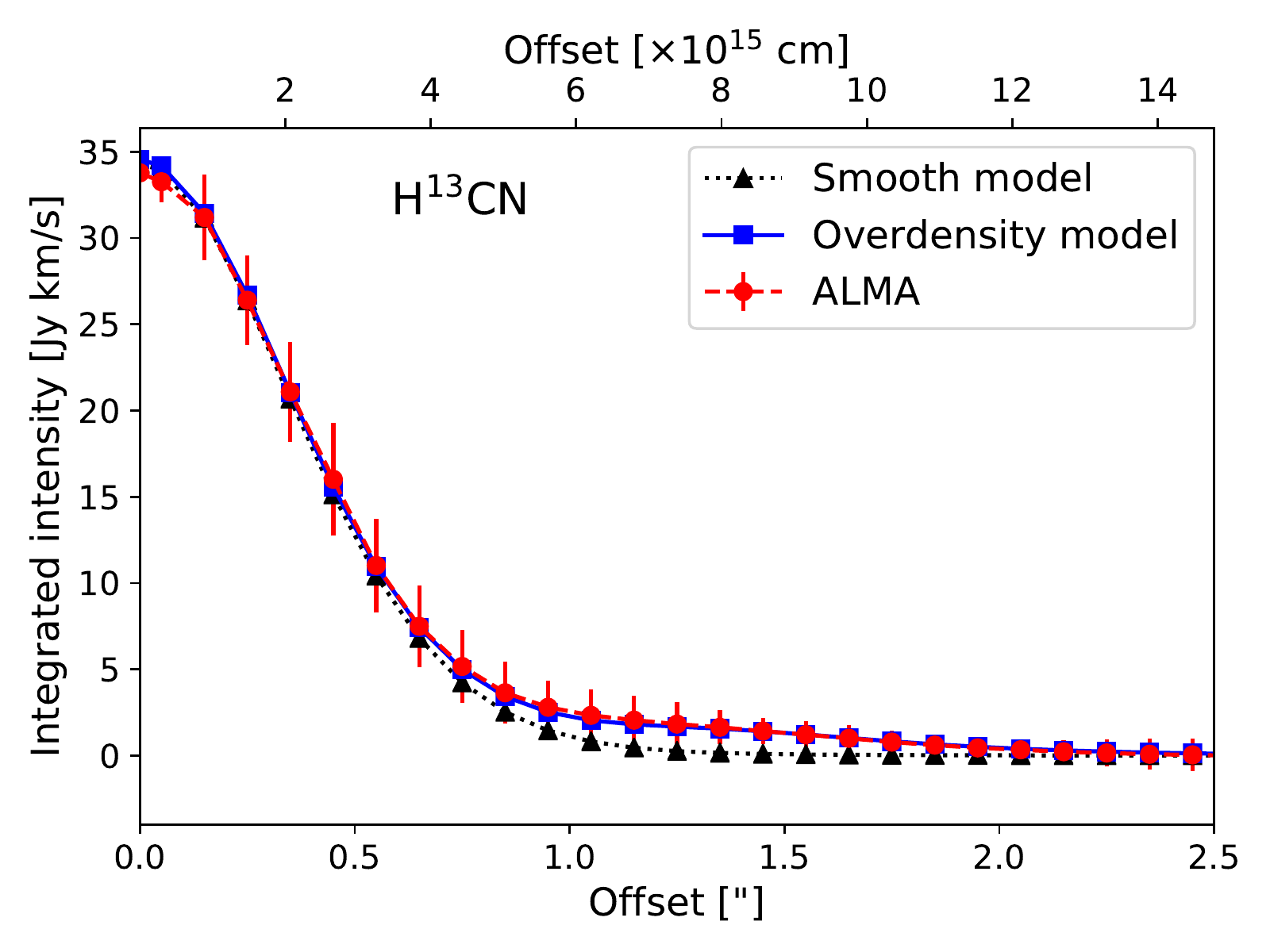} 
   \caption{Radial intensity profiles for the ALMA observations of CS, SiS, $^{30}$SiS, $^{29}$SiO and H$^{13}$CN (red circles and dashed lines) compared to the best fitting radiative transfer model with an overdensity (blue squares and solid lines), and the best fitting smooth-wind radiative transfer model (black triangles and dotted lines).}
   \label{fig:all-positions}
\end{figure*}

A comparison of the radial abundance distributions between all modeled molecules as well as some from the earlier study by \citet{Danilovich2014} is presented in Fig.\,\ref{fig:abundanceprofile}. Similarly, in Fig.\,\ref{fig:all-positions} we plot the best fitting smooth wind and overdensity models against the observed azimuthally averaged radial profiles. There, it is clearly seen that the addition of the overdensity has the most significant impact on the radial profiles of CS and H$^{13}$CN, giving much better fits to the observations from $\sim$1$^{\prime\prime\,}${} outwards.
 Below, we discuss the modeling results for the individual spectral lines. In general, the ALMA observations are reasonably well described by the models when looking at the radial intensity profiles, as are the single-dish observations.

\begin{figure*}[htbp] 
  \centering
   \includegraphics[ width=0.49\textwidth]{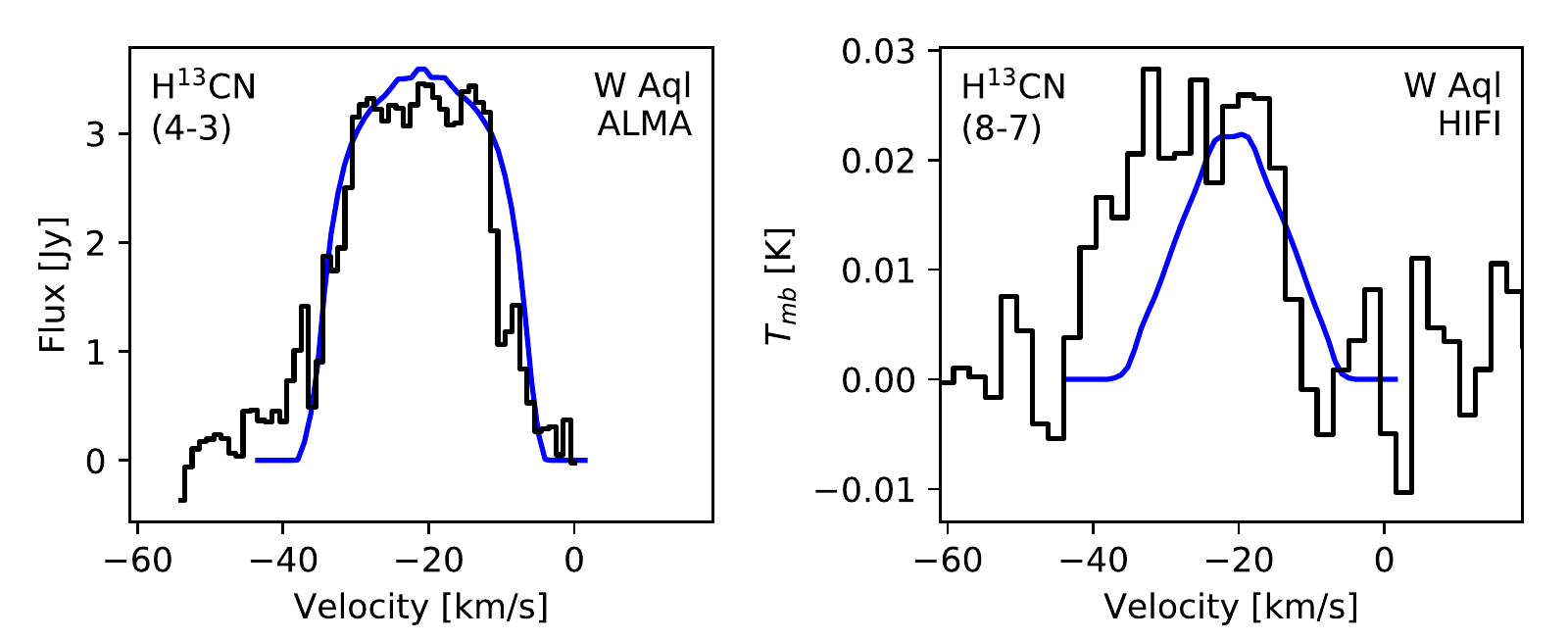} 
   \includegraphics[ width=0.49\textwidth]{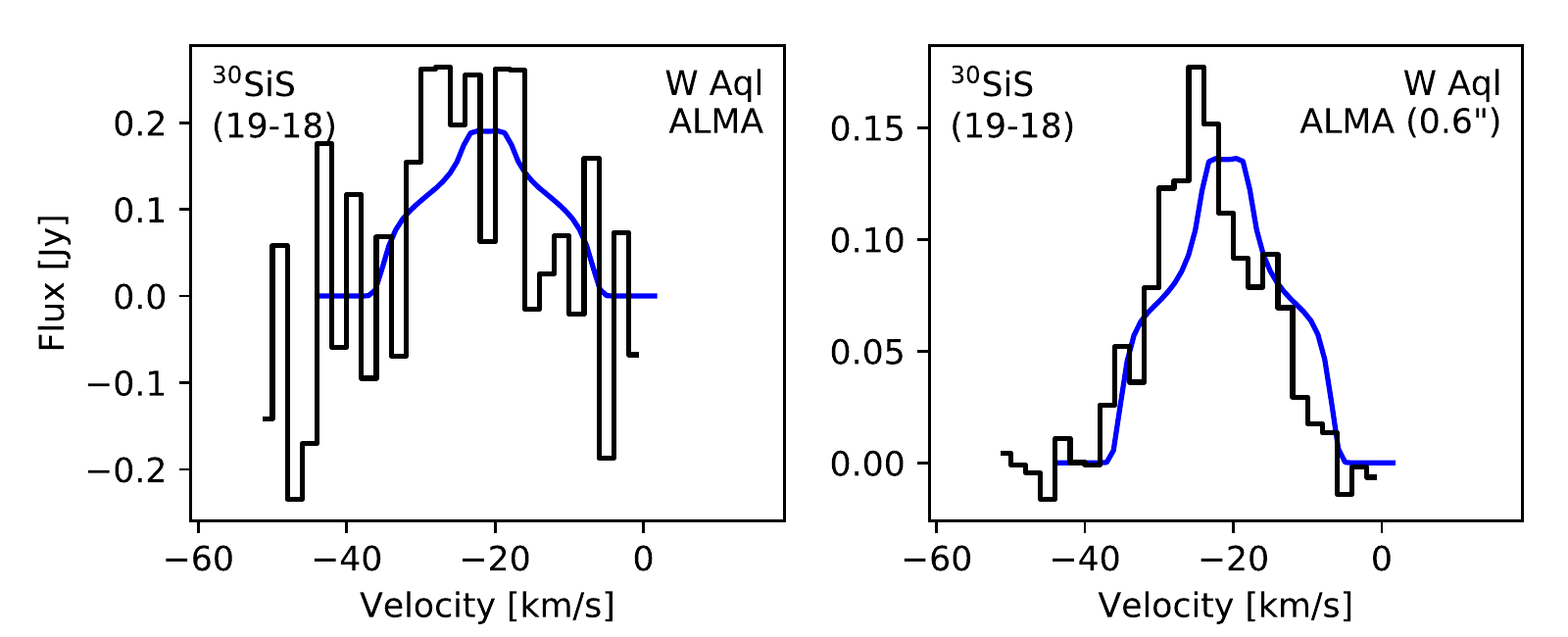}
   \includegraphics[ width=0.49\textwidth]{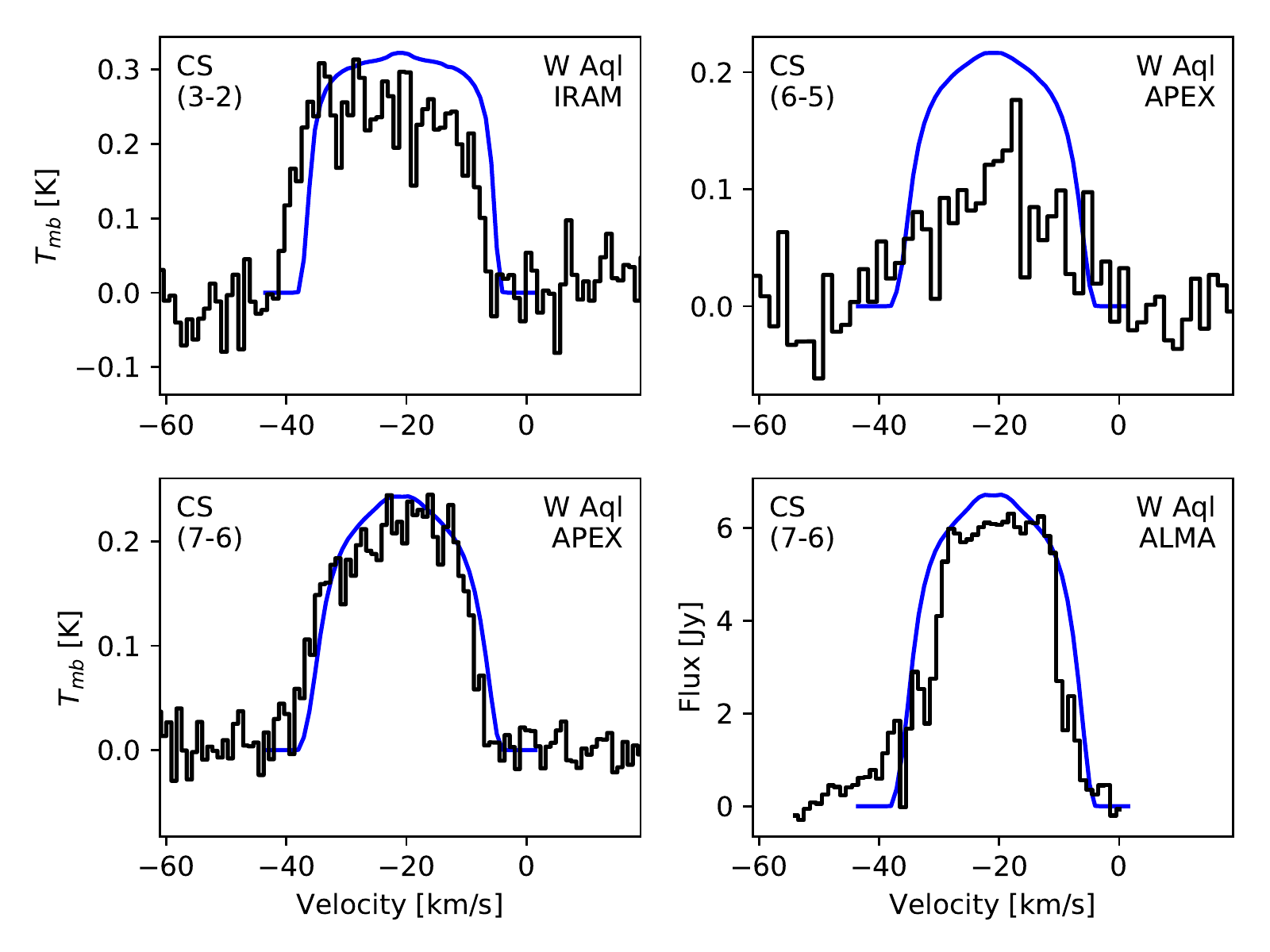}
   \includegraphics[ width=0.49\textwidth]{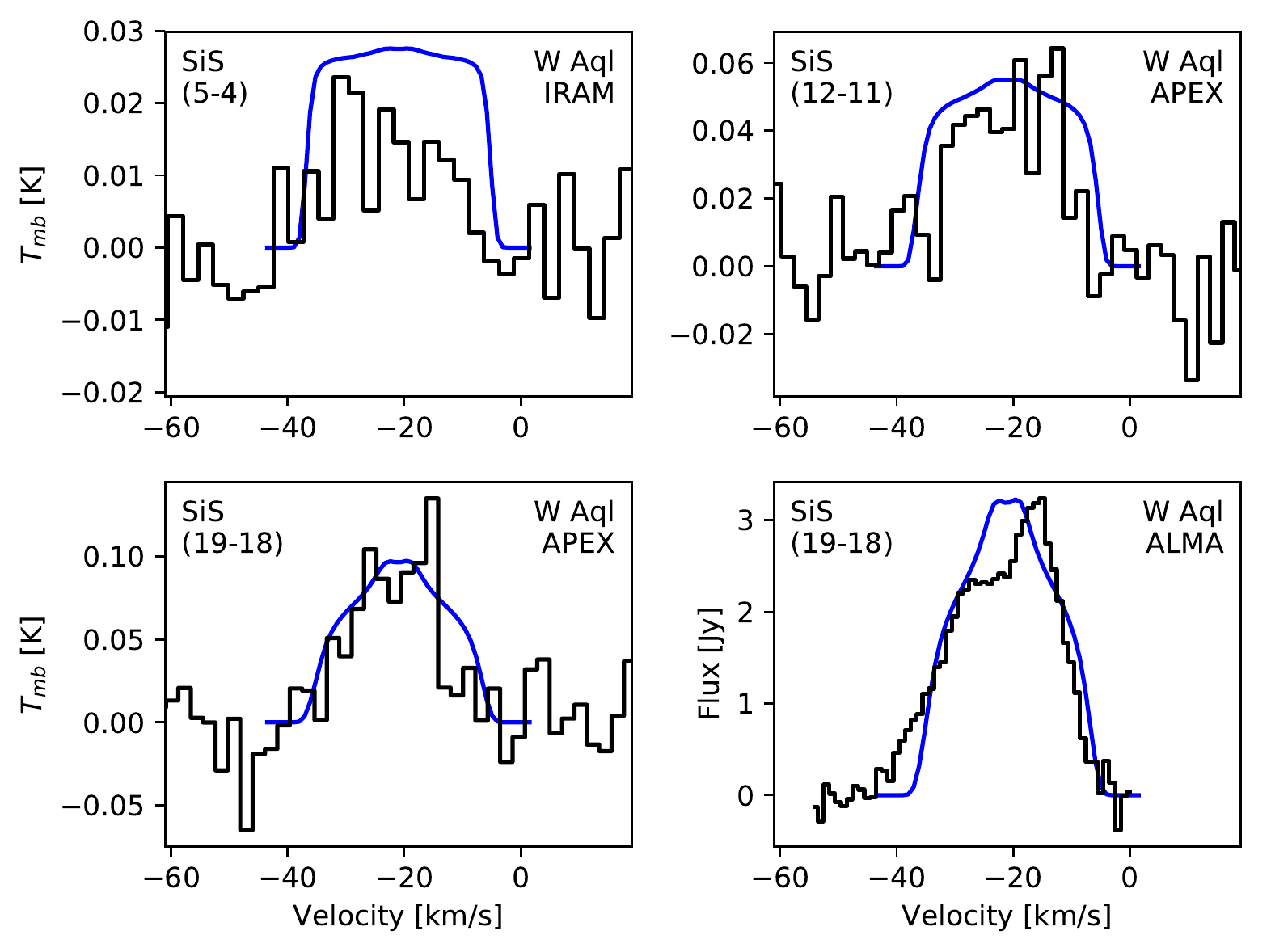}
   \includegraphics[ width=0.49\textwidth]{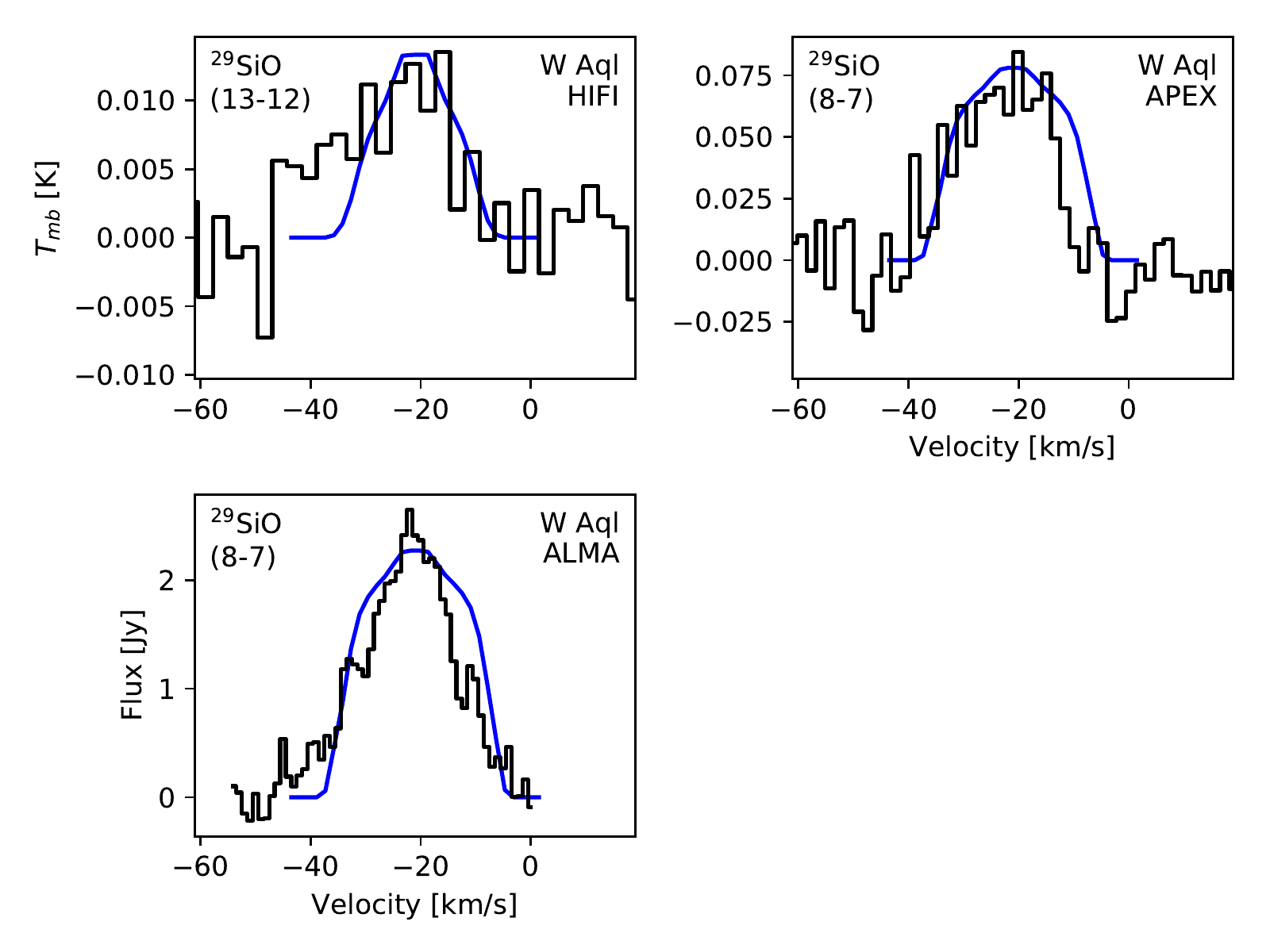}
   \includegraphics[ width=0.49\textwidth]{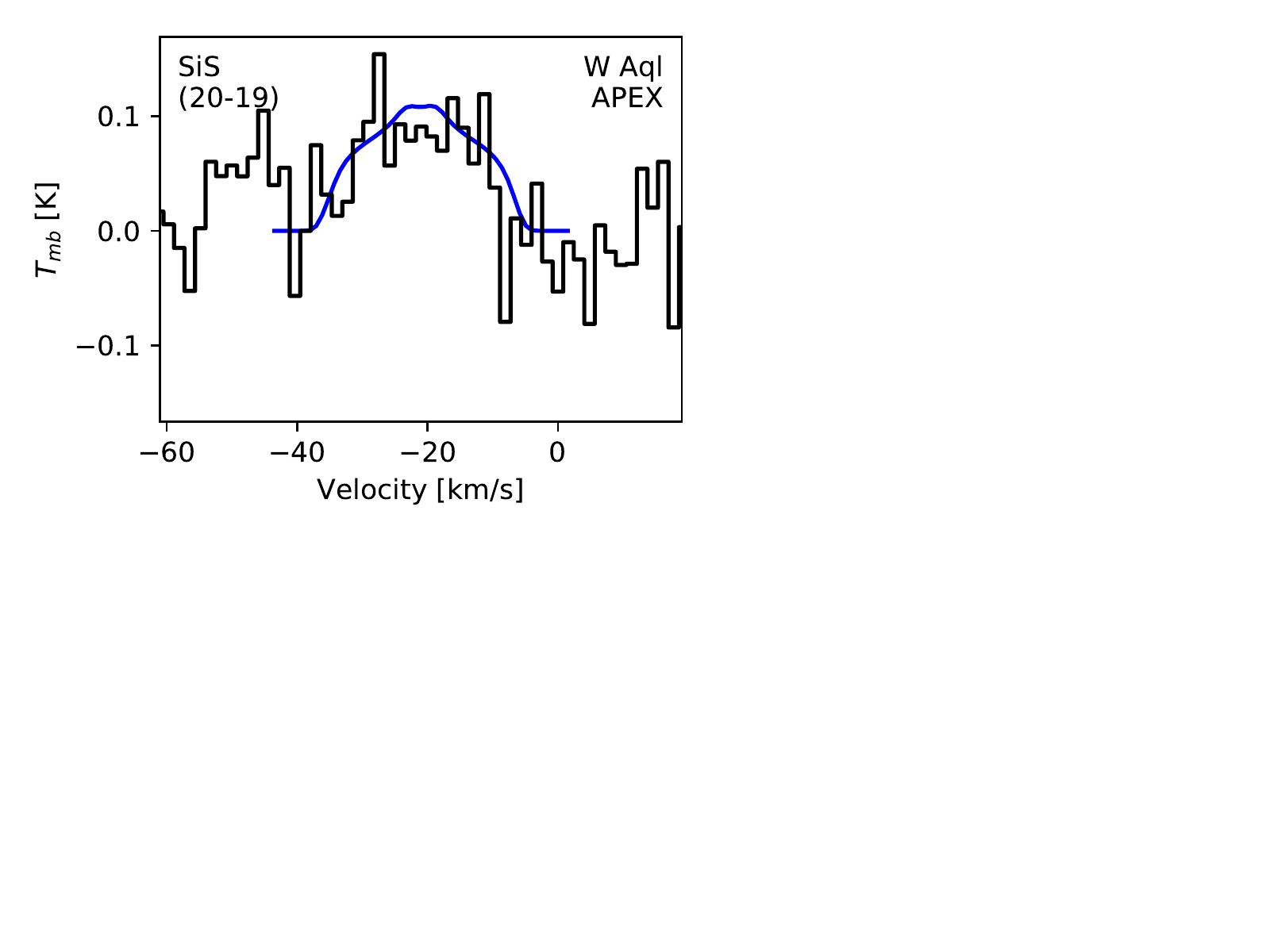}
   \caption{Spectra of different molecular transitions of CS, SiS, $^{30}$SiS, $^{29}$SiO and H$^{13}$CN (black histograms) compared to the best fitting radiative transfer model (blue solid lines).}
   \label{fig:all-spectra}
\end{figure*}


\subsubsection{CS}

For the radiative transfer analysis of CS, we used a molecular data file including the ground- and first excited vibrational state with rotational energy levels from $J = 0$ up to $J=40$, and their corresponding radiative transitions. Our models include the 8$\,\mu m$ CS vibrational band. The energy levels, transition frequencies, and Einstein A coefficients were taken from CDMS \citep{Muller2001,Muller2005} and the collisional rates were adapted from the \citet{Yang2010} rates for CO with H$_2$, assuming an $\mathrm{H}_2$ ortho-to-para ratio of three and scaled to represent collisions between CS and $\mathrm{H}_2$.

For the smooth wind model, we found a fractional abundance for CS of $f_0 = 1.20\times 10^{-6}$  and an $e$-folding radius $R_e = 7.0\times 10^{15}$ cm. While this model fits the inner radial profile and the single dish data well, the outer tail was not well-described by this model. For the overdensity model, using the same Gaussian abundance parameters, we found a much-improved fit, which was improved further by increasing the molecular abundance in the overdense region by a factor of two relative to the Gaussian abundance. The line profile and spectra predicted by the best-fitting (overdensity) model are plotted in Figs.\,\ref{fig:all-positions} and\,\ref{fig:all-spectra}, respectively.

\subsubsection{SiS}

For the radiative transfer analysis of $^{28}$SiS, we used a molecular data file including the rotational levels from $J = 0$ up to $J=40$ for the ground- and first excited vibrational states, and their corresponding radiative transitions. Our models include the 13$\,\mu m$ SiS vibrational band. The energy levels, transition frequencies and Einstein A coefficients were taken from CDMS \citep{Muller2001,Muller2005} and the collisional rates were taken from the \citet{Dayou2006} rates for SiO with He, scaled to represent collisions between SiS and $\mathrm{H}_2$.

For the smooth wind model, we found a fractional abundance for $^{28}$SiS of $f_0 = 1.50\times 10^{-6}$ and an $e$-folding radius $R_e = 6.0\times 10^{15}$ cm. This model agreed with the single-dish observations well, but did not fully agree with the ``tail'' in the azimuthally averaged radial profile from the ALMA observations. We found the model improved with the inclusion of the overdensity and, as with CS, an increase of abundance in the overdense region. However, we found that the overdensity and especially the increase of abundance in the overdense region, had a significant effect on the low-$J$ SiS lines, in a way that was not seen for the low-$J$ CS lines at similar (or lower) energies. This provided an additional constraint on the extent of the overdense region, which we were unable to constrain based solely on the ALMA data, and on the abundance increase in the overdense region. An increase in abundance by a factor of seven in the overdense region provides the best fit when considering only the ALMA observations, but strongly over-predicts the single dish ($5\to4$) and ($12\to11$) SiS lines. Reducing the overdensity abundance down to a factor of three above the smooth wind Gaussian abundance provides the best possible fit when taking both the ALMA radial profile and the single dish observations into account.
The line profile and spectra predicted by the best-fit model are plotted in Figs.\,\ref{fig:all-positions} and\,\ref{fig:all-spectra}, respectively.

For the radiative transfer analysis of $^{30}$SiS, we constructed the molecular data file to contain the equivalent energy levels and transitions as the $^{28}$SiS file. 

The energy levels, transition frequencies and Einstein A coefficients were taken from the JPL molecular spectroscopy database \citep{Pickett1998} and the same collisional rates as for $^{28}$SiS were used, based on rates calculated in \citet{Dayou2006}.

We found a fractional abundance for $^{30}$SiS of $f_0 = 1.15 \times 10^{-7}$ and an $e$-folding radius $R_e = 3.5\times 10^{15}$ cm, which is in general agreement with the $R_e$ found for $^{28}$SiS. The inclusion of the overdensity did not have a significant impact on the $^{30}$SiS model fit to the observations, most likely because the $^{30}$SiS flux is much weaker, especially in the region where the brighter molecules exhibit a ``tail''. As such, we also had no need to increase the $^{30}$SiS abundance in the overdense region. In fact, since the $^{30}$SiS flux is weak compared to the $^{28}$SiS emission at the location of the employed overdensity, testing the $^{28}$SiS overdensity and increased overdensity abundance does not produce a visible change in the $^{30}$SiS model.

The line profile and spectra predicted by the $^{30}$SiS best-fit model are plotted in Figs.\,\ref{fig:all-positions} and\,\ref{fig:all-spectra}. We include a plot of the spectrum at the center of the source convolved with a 0.6$^{\prime\prime\,}${} beam, in addition to the 4$^{\prime\prime\,}${} beam (used for the other molecules), since the 0.6$^{\prime\prime\,}${} beam spectrum has a significantly higher S/N.

\subsubsection{$^{29}$SiO}

We used the same molecular data file as that used for $^{29}$SiO radiative transfer modeling in \citet{Danilovich2014}. This includes the rotational energy levels from $J=0$ up to $J=40$ for the ground and first excited vibrational states. Our models include the 8$\,\mu m$ SiO vibrational band. The collisional rates were also taken from \citet{Dayou2006}.

For the smooth wind model, we found a fractional abundance of $^{29}$SiO of $f_0 = 2.50\times 10^{-7}$ and an $e$-folding radius $R_e = 1.4\times 10^{16}$ cm. The overdensity model provided a better fit to the azimuthally averaged ALMA radial profile, and did not require an additional adjustment to the abundance profile to reproduce the observations.
The line profile and spectra predicted by the best-fit model are plotted in Figs.\,\ref{fig:all-positions} and\,\ref{fig:all-spectra}, respectively.

Our derived $^{29}$SiO $e$-folding radius is in reasonable agreement with the $R_e$ for the $^{28}$SiO model by \citet{Danilovich2014}, indicating that $^{28}$SiO and $^{29}$SiO are co-spatially distributed in the CSE. As an additional test, we re-ran the $^{28}$SiO model using the $R_e$ found for $^{29}$SiO \citep[and the original $f_0 = 2.9\times 10^{-6}$ from][]{Danilovich2014} and found that the smooth wind model was still in good agreement with the single dish $^{28}$SiO data, albeit that the low-$J$ single dish data was somewhat over-predicted by the overdensity model. 

\subsubsection{H$^{13}$CN}

We used the same molecular data file as used for H$^{13}$CN radiative transfer modeling in \citet{Danilovich2014}. This included the rotational energy levels from $J=0$ up to $J=29$ for the ground vibrational state, the CH stretching mode at 3~$\,\mu m$, and the bending mode at 14~$\,\mu m$. The bending mode is further divided into the two $l$-type doubling states. The collisional rates were taken from \cite{Dumouchel2010}.

For the smooth wind model, we found a fractional abundance for H$^{13}$CN of $f_0 = 1.75\times 10^{-7}$ and an $e$-folding radius $R_e = 4.7\times 10^{15}$ cm. H$^{13}$CN was one of the molecules for which the overdensity model made the most significant improvement to the fit of the model. No additional increase in abundance was required in the overdense region to reproduce the ALMA data. We also re-ran the H$^{12}$CN model from \citet{Danilovich2014} with our H$^{13}$CN $R_e$ and the original abundance, $f_0 = 3.1\times 10^{-6}$, and found that the overdense model still reproduced the ($1\to0$) line well. The H$^{13}$CN line profile and spectra are plotted against the model in Figs.\,\ref{fig:all-positions} and\,\ref{fig:all-spectra}, respectively. 

\section{Discussion}
\label{sec:discussion}

\subsection{Comparison with previous studies}
\label{sec:compstudies}

With the ALMA observations at hand, the radial emission size for each measured molecular transition can be extracted directly from the observations and provides an important constraint on the radiative transfer models, which usually use empirical scaling relations to approximate the $e$-folding radii of individual molecules. Additionally, the sensitivity of interferometric observations is superior to single-dish observations, improving the detection threshold and S/N of the observed spectral lines significantly. This also allowed us to determine a model with an overdense region to better represent the ALMA observations --- a feature which cannot be discerned from spatially unresolved observations.

\citet{Danilovich2014} modeled several molecules observed in the CSE of W~Aql with multiple lines available. For $^{29}$SiO and H$^{13}$CN, they only have one available line each and assume the same $e$-folding radii as for the respective main isotopologs. We compare the $e$-folding radius $R_{e}$ for $^{29}$SiO and H$^{13}$CN measured by this study to the calculated $R_{e}$ by \citet{Danilovich2014}, who use the scaling relations by \citet{Gonzalez-Delgado2003b} (for $^{29}$SiO, assuming the same radial distribution as for $^{28}$SiO) and \citet{Schoier2013} (for H$^{13}$CN, assuming the same radial distribution as for H$^{12}$CN) to estimate the $e$-folding radius. 
For $^{29}$SiO, the results roughly agree with each other, where we get a radius of $1.4 \times 10^{16}$ cm and \citet{Danilovich2014} derive a radius of $1.0 \times 10^{16}$ cm. The same is true for the $^{29}$SiO abundance, which \citet{Danilovich2014} find to be $(2.3 \pm 0.6) \times 10^{-7}$, which is in agreement within the errors to our derived abundance of $(2.50 \pm 0.1) \times 10^{-7}$. 
For H$^{13}$CN, we get a radius of $4.7 \times 10^{16}$ cm, while \citet{Danilovich2014} base their model on a radius of $1.8 \times 10^{16}$ cm. Our H$^{13}$CN radius is a factor of approximately three larger than their calculated one. Additionally, we tested the H$^{12}$CN model from \citet{Danilovich2014} (using an abundance of $f_0 = 3.1\times 10^{-6}$) with the $R_e$  for H$^{13}$CN extracted from this study. We found that that model remained in good agreement with both the higher-$J$ lines, such as ($13\to12$), and the low-$J$ lines, including ($1\to0$). The H$^{13}$CN abundance derived by \citet{Danilovich2014} is $(2.8 \pm 0.8) \times 10^{-7}$, which is slightly higher than our result of $(1.75 \pm 0.05) \times 10^{-7}$. Since the model from \citet{Danilovich2014} was based only on one single line that showed a particularly pronounced blue wing excess and the fit was based on the integrated intensity, we expect that this discrepancy is caused by the uncertainty of the single-dish observations.

\citet{Schoier2007} present observations and detailed radiative transfer modeling of SiS for a sample of M-type and carbon AGB stars with varying C/O ratios (no S-type stars included), and assume that the SiS radius would be the same as for SiO, and therefore using the scaling relation of \citet{Gonzalez-Delgado2003b} to calculate the $e$-folding radius. With this in mind, we compare the calculated SiO radius of $1.0 \times 10^{16}$ cm by \citet{Danilovich2014} with our derived SiS radius of $6.0 \times 10^{15}$ cm, which is a factor of 1.7 smaller. Additionally, \citet{Schoier2007} needed to add a compact, highly abundant SiS component close to the star to better fit their models, which we do not require. Indeed our Gaussian abundance distribution fits both the ALMA and single-dish data well, with the overdensity model only required to fit the outer regions of the ALMA radial profile.

 \citet{Decin2008} perform a critical density analysis for SiO, HCN, and CS, which they detect through single-dish observations of a sample of AGB stars -- including W~Aql -- to determine the formation region of the molecular lines. They use two different methods to derive the maximum radius of the emission region (in contrast to the $e$-folding radius): radiative transfer modeling with the GASTRoNOoM code \citep{Decin2006} for collisional excitation, and calculations for radiative excitation only. For CS, they derive $1.68 \times 10^{15}$\,cm and $2.64 \times 10^{14}$\,cm, respectively, using a distance of 230\,pc (with a luminosity of 6800\,L$_{\odot}$ and an effective temperature of 2800\,K). We derive $7.0 \times 10^{15}$\,cm, which is a factor of 4 and 27 larger than their reported values. The different stellar parameters and model setup make it difficult to compare the resulting radii of the emission regions, extracted from different models. As a rough estimate, we scale the reported sizes linearly with the distance and arrive at the conclusion that our CS radius is still bigger than the radii reported by \citet{Decin2008}. Comparing their HCN radii -- $9.6 \times 10^{14}$\,cm and $2.4 \times 10^{14}$\,cm -- to our measured H$^{13}$CN radius of $4.7 \times 10^{16}$\,cm, we get a substantially larger radius, which also holds for the linear distance scaling. The same is true for the comparison of their SiO radii -- $9.6 \times 10^{14}$\,cm and $2.16 \times 10^{14}$\,cm -- to our measured $^{29}$SiO radius of $1.4 \times 10^{16}$\,cm.

\subsection{Limitations of radiative transfer models}

The most obvious limitation of the radiative transfer models is that they assume spherical symmetry, which, as we can see from the extended emission in the channel maps, is clearly not a precise description of the more complex emission. This issue becomes more important for the strongest spectral lines, those of CS and H$^{13}$CN, where the asymmetric features are more pronounced in the observations. For those lines, we see a clear difference between the smooth wind models and the observed radial intensity profiles at distances greater than 1$^{\prime\prime\,}$ from the star (Fig.\,\ref{fig:all-positions}). This is also visible in a much weaker form for the SiS and $^{29}$SiO lines (also Fig.\,\ref{fig:all-positions}). Nevertheless, \citet{Ramstedt2017} show that the clearly visible and pronounced arc structures in the CO density distribution are not very pronounced in the overall radial density profile, but are simply represented by rather weak features on top of a smooth extended component. Here, in our overdensity model, we add a higher-density component to the smooth wind model (see Fig. \ref{fig:nh2plot}) in a region that matches the ALMA radial profiles. Although the actual overdense region is unlikely to be spherically symmetric in reality \citep[as the arcs seen in the CO emission by][are not]{Ramstedt2017}, we find it to be well fit by our spherically symmetric model.

Another limitation is the direct comparison of often noisy and low-resolution single-dish spectra (often subject to uncertain calibration) with high S/N and high-resolution ALMA spectra as model input. The CS single-dish data in particular generally has a low S/N, and two of the single-dish spectra are not well fitted (Fig.\,\ref{fig:all-spectra}). Since the single-dish data were taken at many different epochs, the stellar variability might have an influence on the extracted spectra as well \citep[e.g.,][]{Teyssier2015,Cernicharo2014}
Additionally, since we are probing high-spatial-resolution asymmetric and clumpy emission with the ALMA observations, the line profiles from the symmetric radiative transfer models are unlikely to be completely representative of the real complexity of the CSE around W\,Aql and therefore do not fit the observations perfectly. Nevertheless, the general agreement of the models with observations is good.

\subsection{Comparison with chemical models}

From the theoretical models of \citet{Cherchneff2006} and a sample of observations by \citet{Bujarrabal1994} and \citet{Schoier2007}, it is evident that the SiS abundance is sensitive to the chemical type of AGB stars, with carbon stars showing generally higher abundances than M-type stars. The mean fractional SiS abundances of carbon and M-type stars reported by \citet{Schoier2007} are $3.1 \times 10^{-6}$  and $2.7 \times 10^{-7}$, respectively. Our derived SiS abundance of $1.5\times 10^{-6}$ for W\,Aql as an S-type star falls close to the lower boundary of the expected region of carbon AGB stars \citep[compared with Fig.\,2 of][]{Schoier2007}.
Predicted by the non-LTE chemical models of \citet{Cherchneff2006}, CS should be found in high abundance for all types of AGB stars in the inner wind, with a generally higher abundance of CS for carbon stars than M-type stars. The formation of CS is linked to HCN, which is predicted to be present in comparable abundances. As discussed by \citet{Cherchneff2006}, in their models, the maximum abundance of HCN -- and therefore CS -- is reached for S-type AGB stars. In Fig.\,\ref{fig:abundanceprofile}, one can see that, for W~Aql, the radial abundance profile of HCN \citep[as derived by][]{Danilovich2014} is indeed comparable to the profile of CS (excluding the overdensity enhancement), which is generally of slightly lower abundance and extends out to a slightly smaller radius. The abundance ratio of those molecules is HCN/CS $= 3.1 \times 10^{-6} / 1.3 \times 10^{-6} \simeq 2.4$. In the theoretical models of \citet{Cherchneff2006}, the HCN/CS ratio for a ``typical'' AGB with a C/O ratio between 0.75 and 1 lies between 5.34 and 0.28, at radii larger than 5\,R$_{\star}$. The best fitting theoretical HCN/CS ratio to our modeled ratio of 2.4 would be reached for a C/O ratio somewhere between 0.90 and 0.98. The theoretical HCN/CS ratio of 0.28 for a C/O ratio of 1 does not fit our models.  A very similar value can be derived from thermodynamic equilibrium calculations (Gobrecht, priv. comm.), which yield a HCN abundance of $4.31 \times 10^{-8}$ and a CS abundance of $1.38 \times 10^{-7}$, resulting in a HCN/CS ratio of 0.31. It must be noted, however, that the chemical models are focused on the innermost winds and our observations detect molecular emission as far away from the star as $\sim$4$^{\prime\prime\,}$. Additionally, the stellar temperature can affect the chemical equilibrium and therefore the chemical abundances.

On a general note on sulphur bearing molecules in CSEs of AGB stars, \citet{Danilovich2016} found a very different S chemistry between low mass-loss rate and high mass-loss rate M-type stars when examining SO and SO$_2$, which was not predicted by chemical models.

\subsection{Isotopic ratios}
\subsubsection{Silicon}
Previous models by \citet{Danilovich2014} estimate a $^{28}$SiO abundance of $2.9 \times 10^{-6}$, which we can use to derive the $^{28}$SiO/$^{29}$SiO ratio, resulting in a value of 11.6. From our model estimates, we find the abundance ratio of $^{28}$SiS/$^{30}$SiS =13.0, roughly corresponding to half of the $^{28}$Si/$^{30}$Si ratio predicted by AGB stellar evolution models \citep{Cristallo2015, Karakas2010} and half of the value given by the solar abundances \citep{Asplund2009}. By assuming that both SiO and SiS ratios are good tracers for the Si ratios, we can approximate the $^{29}$Si/$^{30}$Si ratio for W\,Aql, which results in a value of 1.1. For comparison, the solar $^{29}$Si/$^{30}$Si ratio, as derived by \citet{Asplund2009} is 1.52, which is also consistent with the ratio derived for the ISM \citep{Wolff1980,Penzias1981}. \citet{Peng2013} investigate the $^{29}$Si/$^{30}$Si ratio for 15 evolved stars and come to the conclusion that the older low-mass oxygen-rich stars in their sample have lower $^{29}$Si/$^{30}$Si ratios, but the ratios are not influenced strongly by AGB evolution and merely reflect the interstellar environment in which the star has been born. The only S-type star in their sample, $\chi$\,Cyg, shows a $^{29}$Si/$^{30}$Si ratio of 1.1, which is incidentally identical to our derived value for W~Aql. A bigger sample of Si isotopic ratios derived for S-type stars would be needed to draw any conclusions on the possible AGB evolutionary effects.

\subsubsection{Carbon}
As described by \citet{Saberi2017} for the carbon AGB star R\,Scl, the H$^{12}$CN/H$^{13}$CN ratio is a very good tracer for the $^{12}$C/$^{13}$C ratio, giving even more plausible results than estimates from the CO isotopic ratio in some cases. Using modeling results by \citet{Danilovich2014} for H$^{12}$CN, which has an abundance of $(3.1 \pm 0.1) \times 10^{-6}$, and the H$^{13}$CN abundance presented in this paper, we derive a H$^{12}$CN/H$^{13}$CN ratio of $17 \pm 6$ (with the error mainly dominated by the H$^{12}$CN model), which can be used to estimate the $^{12}$C/$^{13}$C ratio. This result differs by a factor of 1.7 to 1.5 from the $^{12}$C/$^{13}$C ratio estimated from the CO isotopolog ratio by \citet{Danilovich2014} and \citet{Ramstedt2014}, who report a ratio of 29 and 26, respectively.  \citet{Ramstedt2014} also present C isotopic ratios for 16 other S-type AGB stars (in addition to 19 C- and 19 M-type AGB stars) and the resulting median carbon isotopic ratio for S-type stars is reported to be 26, which is also their derived value for W\,Aql. In general, the $^{12}$C/$^{13}$C ratio increases following the evolution from M- to S- to C-type stars, since $^{12}$C is primarily produced in the He burning shell of the star and will be dredged up to the surface through subsequent thermal pulses. Therefore, the $^{12}$C/$^{13}$C ratio can be used as an indicator for the AGB evolutionary stage of a star.

\section{Conclusions}
\label{sec:conclusions}

We present ALMA observations of a total of eight molecular lines of the species CS, SiS, SiO and H$^{13}$CN, which are found in the close CSE of the S-type binary AGB star W~Aql. The emission is resolved for almost all lines and a compact Gaussian emission component is surrounded by weak asymmetric emission, showing an elongation in the direction of the binary orbit. Adding an additional delta function emission component to the models at the same position as the Gaussian significantly improves the fits, carried out in the uv plane. Emission peaks in molecular emission out to approximately 2--3$^{\prime\prime\,}$ coincide with the CO distribution. The spectral lines -- especially the strongest four -- are partly very asymmetric, with SiS standing out of the sample with a very distinct peak at the red-shifted velocities.
We successfully model the circumstellar abundance and radial emission size of CS, SiS, $^{30}$SiS, $^{29}$SiO and H$^{13}$CN -- mainly based on high-resolution sub-millimeter observations by ALMA and supplemented by archival single-dish observations -- in the innermost few arcseconds of the CSE around the S-type AGB star W~Aql. We produce two models, one assuming a smooth outflow, and one which approximates the asymmetries seen in the outflow by the inclusion of an overdense component, and which better fits the ALMA data. We compare the resultant radial abundance distributions to previous modeling and theoretical predictions, and find that our results are in agreement with previous studies. For some isotopologs, such as  $^{29}$SiO and H$^{13}$CN, previous studies worked with the assumption of a radial distribution identical to their main isotopologs,   $^{28}$SiO and H$^{12}$CN. We find that our modeled radial distribution of the $^{29}$SiO and H$^{13}$CN isotopologs, which is constrained by the resolved ALMA observations, is different to the main isotopologs, with both rarer isotopologs having a larger radius than previously found for the most common isotopolog counterparts.

By using the spatially resolved ALMA data, we can constrain the radial emission profile directly, which improves the independence of our models by reducing the number of input assumptions. Additionally, we need to develop improved models, which can take asymmetries into account. This and future high-resolution high sensitivity spectral scans of CSEs around AGB stars will provide more accurate input to chemical modeling and help us understand the chemical network of elements in these environments.

\begin{acknowledgements}
This paper makes use of the following ALMA data: ADS/JAO.ALMA$\#$2012.1.00524.S. ALMA is a partnership of ESO (representing its member states), NSF (USA) and NINS (Japan), together with NRC (Canada), NSC and ASIAA (Taiwan), and KASI (Republic of Korea), in cooperation with the Republic of Chile. The Joint ALMA Observatory is operated by ESO, AUI/NRAO and NAOJ. We want to thank the Nordic ARC Node for valuable support concerning ALMA data reduction and uv-fitting, and David Gobrecht for his input to the chemical discussion. MB acknowledges the support by the uni:docs fellowship and the dissertation completion fellowship of the University of Vienna. EDB is supported by the Swedish National Space Board. TD acknowledges support from the Fund of Scientific Research Flanders (FWO).

\end{acknowledgements}

%
%

\bibliographystyle{aa}
\bibliography{FULL-literature}


\begin{appendix} 

\section{Channel maps for strong spectral lines}
\label{sec:appendix1}

   \begin{figure*}
   \centering
   \includegraphics[trim={10cm 9cm 4.5cm 9cm},clip,width=0.98\textwidth]{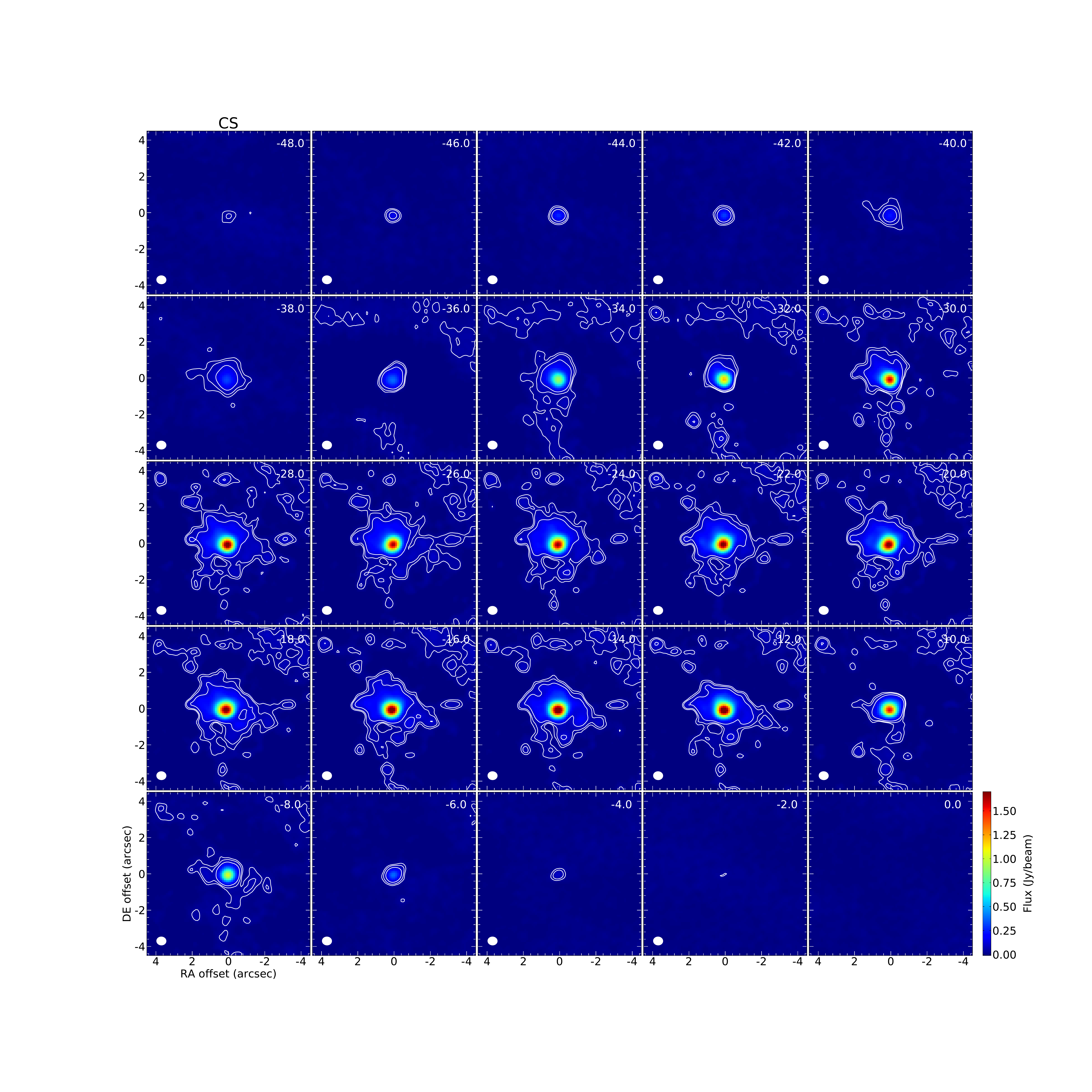}
   \caption{Channel map for the CS spectral line. The velocity resolution is 1\,km~s$^{-1}$, only every second channel is plotted. The stellar velocity is at about -21\,km~s$^{-1}$. The beam size is given in the lower left and the white contours are plotted for 3, 5, and 10\,$\sigma$ rms (measured from line free channels). North is up and east is left.}
              \label{fig:CSmap}%
    \end{figure*}

   \begin{figure*}
   \centering
   \includegraphics[trim={10cm 9cm 4.5cm 9cm},clip,width=0.98\textwidth]{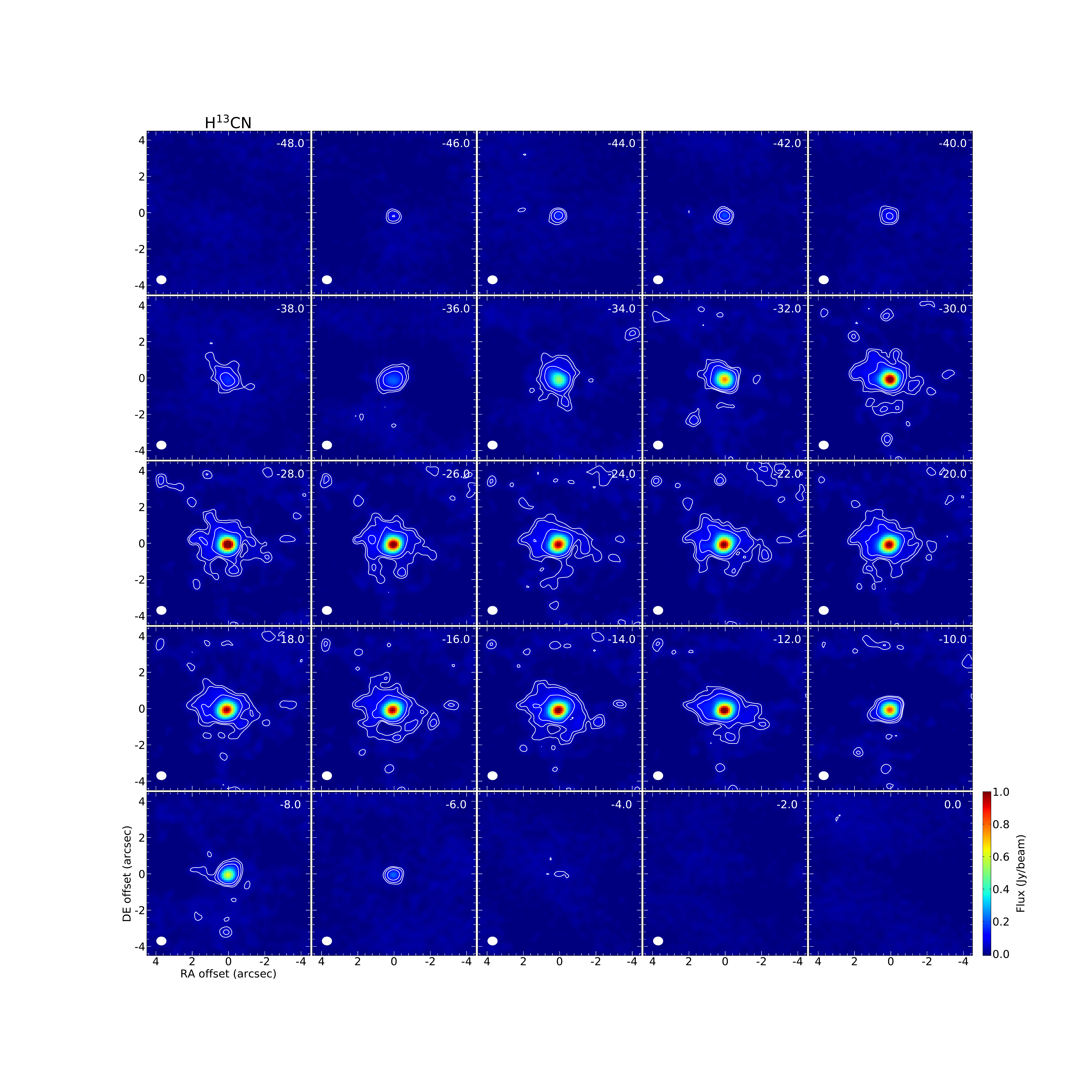}
   \caption{Channel map for the H$^{13}$CN spectral line. The velocity resolution is 1\,km~s$^{-1}$, only every second channel is plotted. The stellar velocity is at about -21\,km~s$^{-1}$. The beam size is given in the lower left and the white contours are plotted for 3, 5, and 10\,$\sigma$ rms (measured from line free channels). North is up and east is left.}
              \label{fig:H13CNmap}%
    \end{figure*}

       \begin{figure*}
   \centering
   \includegraphics[trim={10cm 9cm 4.5cm 9cm},clip,width=0.98\textwidth]{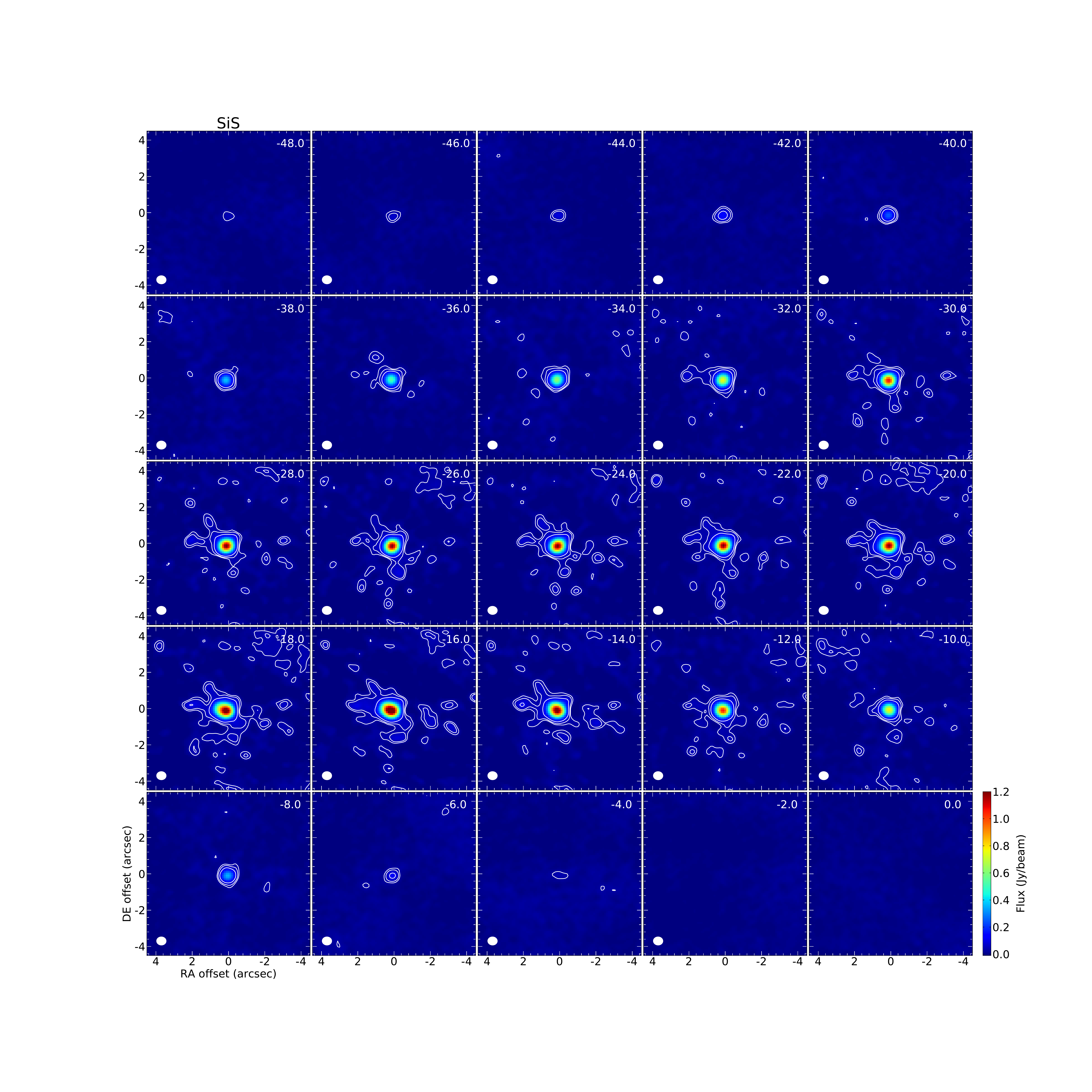}
   \caption{Channel map for the SiS spectral line. The velocity resolution is 1\,km~s$^{-1}$, only every second channel is plotted. The stellar velocity is at about -21\,km~s$^{-1}$. The beam size is given in the lower left and the white contours are plotted for 3, 5, and 10\,$\sigma$ rms (measured from line free channels). North is up and east is left.}
              \label{fig:SiSmap}%
    \end{figure*}

       \begin{figure*}
   \centering
   \includegraphics[trim={10cm 9cm 4.5cm 9cm},clip,width=0.98\textwidth]{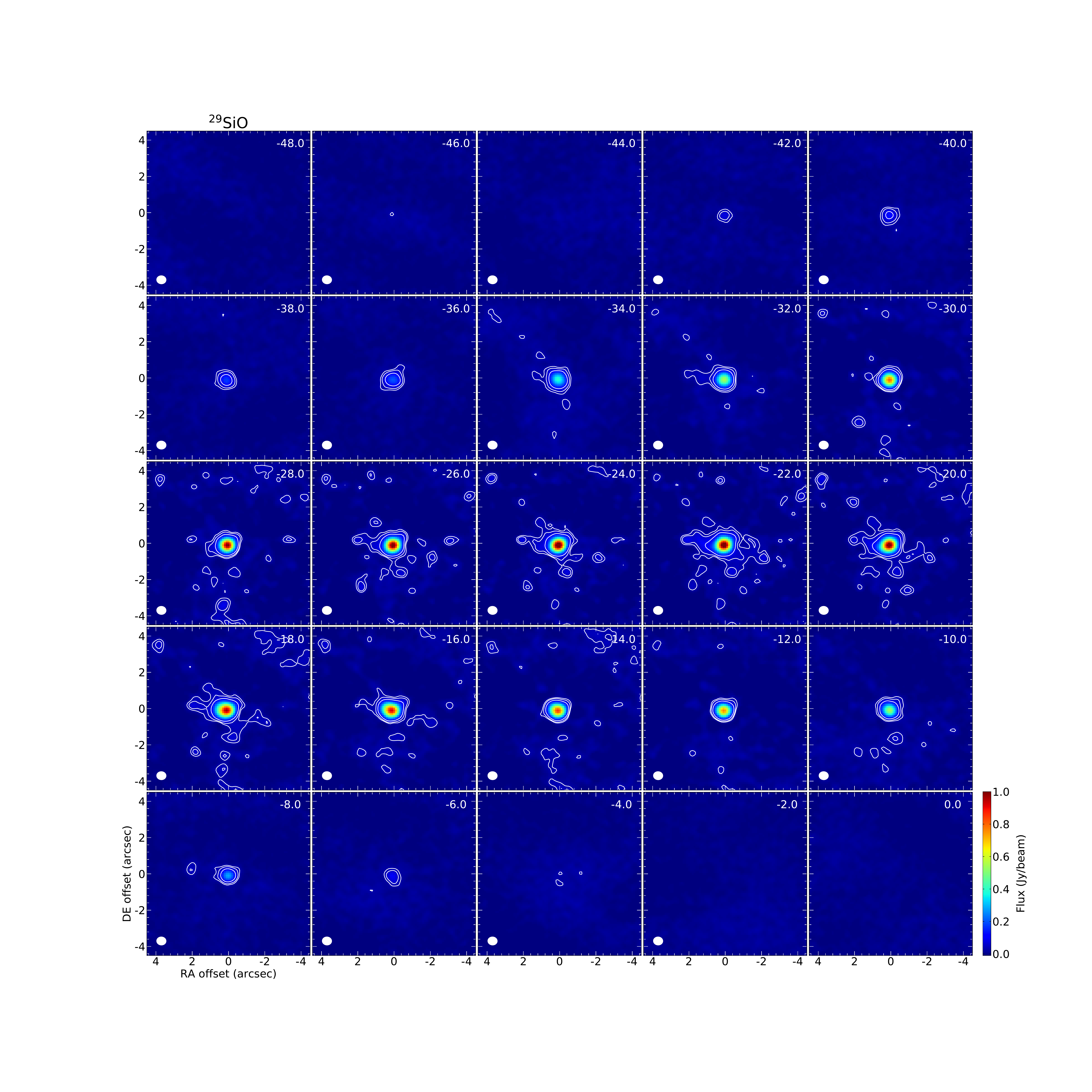}
   \caption{Channel map for the $^{29}$SiO spectral line. The velocity resolution is 1\,km~s$^{-1}$, only every second channel is plotted. The stellar velocity is at about -21\,km~s$^{-1}$. The beam size is given in the lower left and the white contours are plotted for 3, 5, and 10\,$\sigma$ rms (measured from line free channels). North is up and east is left.}
              \label{fig:29SiOmap}%
    \end{figure*}

\section{Results for weak spectral lines}
\label{sec:appendix2}

   \begin{figure*}
   \centering
   \includegraphics[trim={7.5cm 0cm 7.5cm 3cm},clip,width=0.98\textwidth]{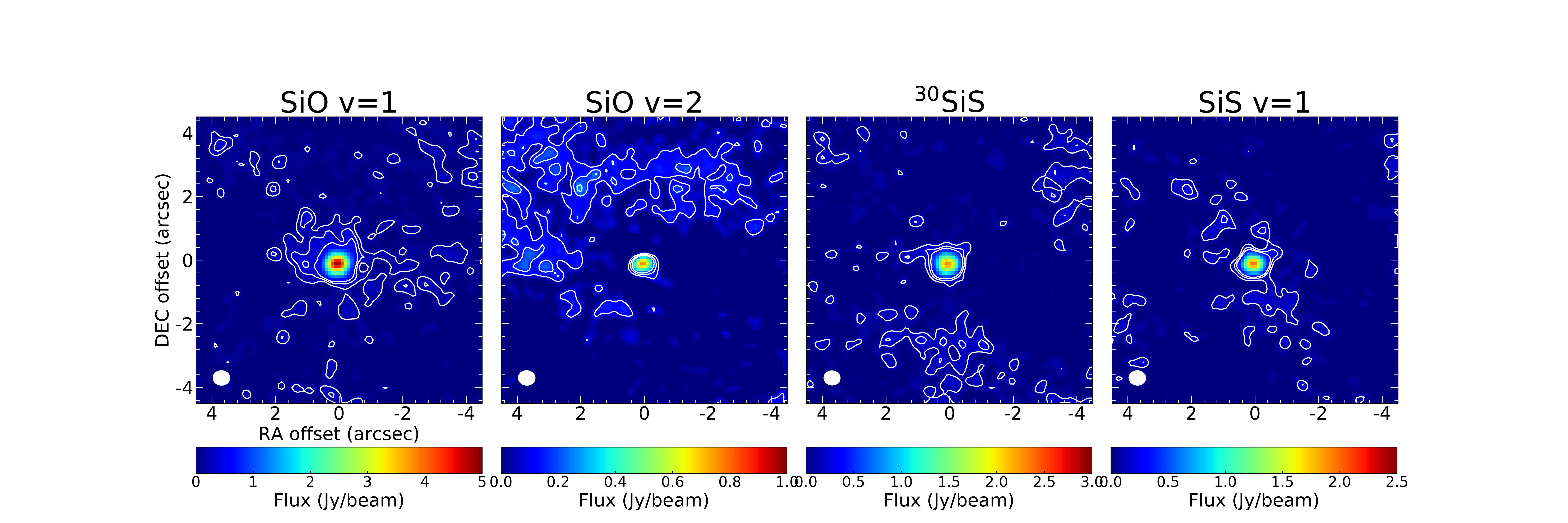}
   \caption{ALMA integrated intensity (moment 0) maps of SiO\,v=1\,(8--7), SiO\,v=2\,(8--7), $^{30}$SiS\,(19--18) and SiS\,v=1\,(19--18) emission around W\,Aql. The ALMA beam is given in the lower-left corner. Contours are given in white for 1, 2, and 3\,$\sigma$ rms. North is up and east is left.}
              \label{fig:weakmoments}%
    \end{figure*}



\section{uv-fitting: super-resolution and uncertainties}
\label{appendixuv}

As has been discussed in \citet{Marti-Vidal2012}, high S/N interferometric observations may encode information about source substructures (much) smaller than the diffraction limit of the interferometer. Extracting such information, though, implies the use of \textit{a priori} assumptions related to the source brightness distribution (e.g., the assumption of a Gaussian intensity distribution that we are using in our modeling). As long as the true source structure obeys these \textit{a priori} assumptions (at least to the extent of the structure sizes being probed in the visibility analysis), it is possible to retrieve structure information that roughly scales as the diffraction limit divided by the square root of the S/N \citep[see, e.g., Eq. 7 of ][]{Marti-Vidal2012}. We notice, though, that this is an approximate rule of thumb that may be corrected by a factor of several, since the actual over-resolution power may depend on the baseline distribution of the array \citep[i.e., the factor $\beta$ in Eq. 7 of ][]{Marti-Vidal2012}.

The modeling performed by \textit{uvmultifit} \citep{Marti-Vidal2014} estimates the uncertainties of the fitted parameters from the curvature of the $\chi^2$ distribution at its minimum via the diagonal elements of the post-fit covariance matrix. The uncertainties provided in this way assume that the model used in the fit \textit{is correct and able to describe the bulk of the visibility signal}, meaning that the reduced $\chi^2$ can be assumed to be similar to its expected value of unity.

\begin{figure*}
   \centering
   \includegraphics[trim={0cm 0cm 0cm 0cm},clip,width=0.98\textwidth]{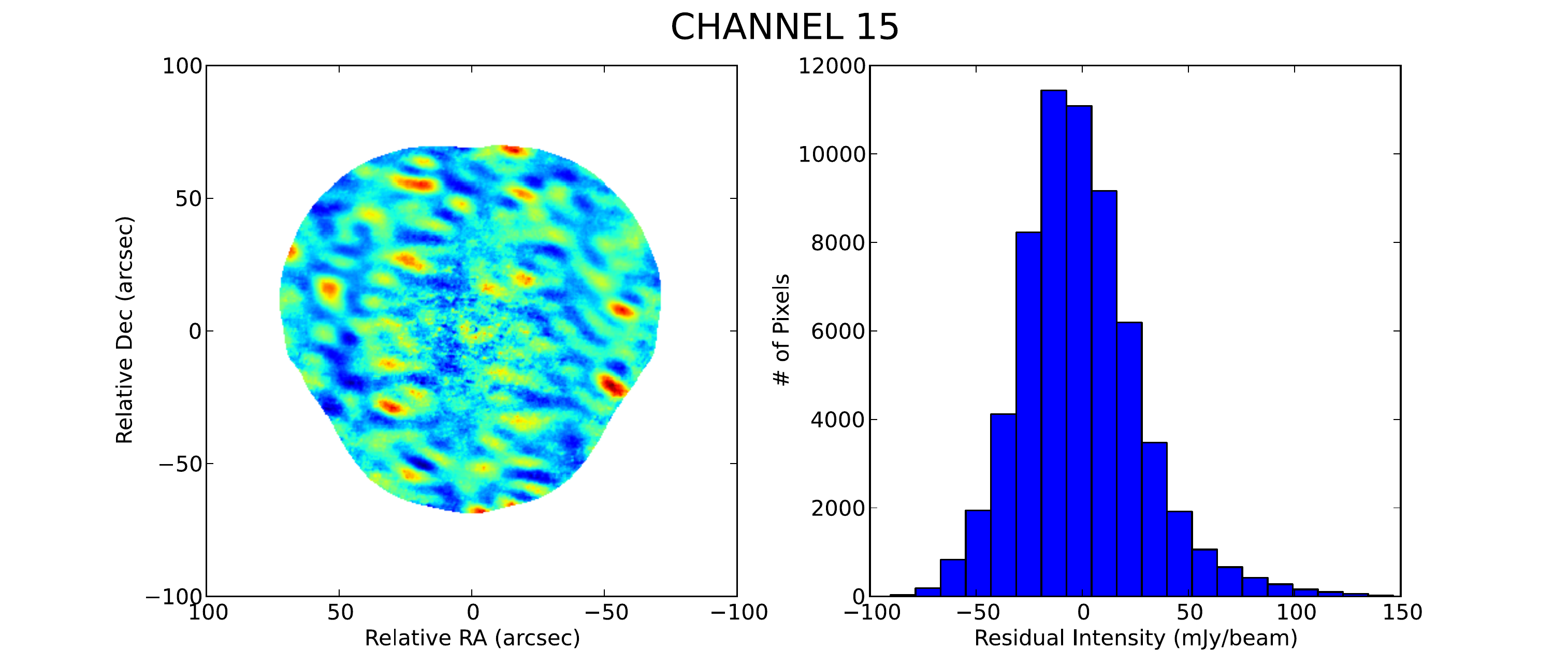}
   \caption{Left: Post-fit residual image for channel 15 (stellar velocity), using natural weighting (i.e., to maximise the detection sensitivity for any flux-density contribution left in the fitting). Right: Histogram of the post-fit image residuals.}
              \label{fig:residuals}%
    \end{figure*}

This assumption may result in underestimates of the parameter uncertainties \textit{if} the model used in the fitting is incorrect. One way to test whether our fitting model is able to describe the bulk of the visibility function is to assess the post-fit residuals in the image plane. We show these residuals for a selection of two channels in the SiS line in Fig.\,\ref{fig:residuals} (left). The intensity distribution of the residuals can be well characterized by Gaussian noise (Fig.\,\ref{fig:residuals}, right), which is indicative of a successful fit.

\begin{figure*}
   \centering
   \includegraphics[trim={0cm 0cm 0cm 1.25cm},clip,width=0.98\textwidth]{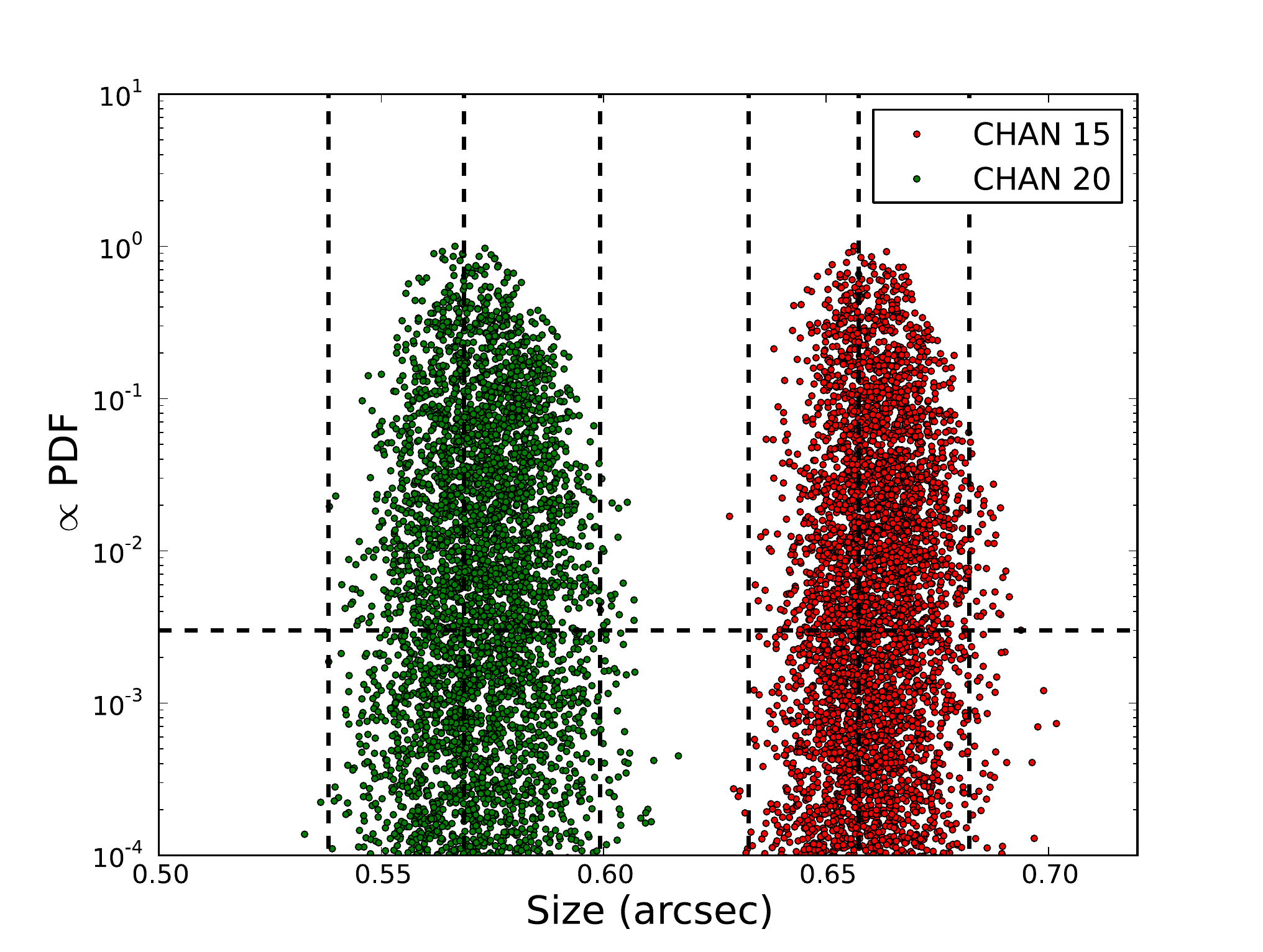}
   \caption{Probability density distributions from our model-fitting for the SiS line of channels 15 and 20, using a Gaussian source model with addition of a delta function model. The vertical dashed lines indicate the estimated best-fit values and their $\pm 3\sigma$ deviations. The horizontal dashed line is set to $3\times10^{-3}$ (i.e., the probability density corresponding to the $3\sigma$ cutoff of a Gaussian probability distribution). The peaks of the PDFs have been further scaled to 1, for clarity.}
              \label{fig:mcsim}%
    \end{figure*}

In order to further assess our size estimates and uncertainties, we have performed a Monte Carlo analysis of our Gaussian source model plus additional delta component across the full parameter space (i.e., varying the peak position, flux density, and size with a random exploration), to test how the $\chi^2$ (and hence the probability density) varies in the neighborhood of the best-fit parameter values. We show the resulting \textit{probability density functions} (PDF) for channels 13 and 20 in Fig.\,\ref{fig:mcsim}. We notice that the standard assumptions of \textit{uvmultifit} are applied here (i.e., a reduced $\chi^2$ of 1 is assumed, which is justified by the post-fit residuals; see Fig.\,\ref{fig:residuals}). The crossing points between the 3$\sigma$ deviations of the source size (vertical dashed lines) and the 3$\sigma$ probability cutoff for a Gaussian PDF (horizontal dashed line) roughly coincide with the crossing points between the 3$\sigma$ source-size deviation and the PDF retrieved with \textit{uvmultifit}.

We can therefore conclude that, as long as the Gaussian plus delta component brightness distribution is a good approximation of the source structure, the source sizes (and their uncertainties) obtained with \textit{uvmultifit} are correct. If the source shape were to not resemble a Gaussian brightness distribution, but were closer to, say, a disk or a filled sphere, there would be a \textit{global systematic bias} in all source sizes reported here, even though the relative values among frequency channels (i.e., the frequency dependence of the source sizes) would still be correct.

\section{List of archival spectral line observations} 
\label{appendix2}

\begin{table*}[tp]
\caption{Archival observations.}
\label{archivalobs}
\begin{center}
\begin{tabular}{lccccccl}
\toprule
Molecule                & Transition            &Frequency&     Telescope& $\theta$&  $I_{mb}$ & Reference\\
&&[GHz]&&[$^{\prime\prime\,}$]&[K \,km~s$^{-1}$]\\
\midrule
CS      &       $       3 \to 2 $       &       146.969 &       IRAM    &       17      & 6.96 & \citet[][submitted]{Danilovich2018}\\
 & $6\to5$ & 293.912 & APEX & 21 & 2.73 & APEX Archive\\
 & $7\to6$ & 342.883 & APEX & 18 &5.69 & APEX Archive\\
SiS     &       $       5 \to 4 $       &       \phantom{1}90.772       &       IRAM    &       27      & 0.35 & \citet[][submitted]{Danilovich2018}\\
 & $12\to11$ & 217.818& APEX& 29 & 3.37 & \citet[][submitted]{Danilovich2018}\\
 & $19\to18$ & 344.779 & APEX& 18 &1.58 & \citet[][submitted]{Danilovich2018}\\
 & $20\to19$ & 362.907 & APEX& 17 &2.5 & APEX Archive\\
$^{29}$SiO & $13\to12$ & 557.179 & HIFI & 38 & 0.25 & \citet{Danilovich2014}\\
 & $8\to7$ & 344.226 & APEX & 18 & 1.35 & APEX Archive\\
H$^{13}$CN & $8\to7$ & 1151.452\phantom{1} & HIFI &  18.4 & 0.63 & \citet{Danilovich2014}\\

\bottomrule
\end{tabular}
\end{center}
\end{table*}%

\end{appendix}

\end{document}